\documentclass[twocolumn,twocolappendix]{aastex631}
\usepackage{newtxtext,newtxmath,amssymb}
\usepackage[normalem]{ulem}

\renewcommand{\vec}[1]{\pmb{#1}}

\usepackage{calligra}

\begin{document}


\title{Three-Dimensional Numerical Simulations of Magnetar Crust Quakes}

\author[0000-0003-4721-4869]{Yuanhong Qu}\thanks{E-mail: yuanhong.qu@unlv.edu}
\affiliation{Nevada Center for Astrophysics, University of Nevada, Las Vegas, NV 89154}
\affiliation{Department of Physics and Astronomy, University of Nevada Las Vegas, Las Vegas, NV 89154, USA}

\author[0000-0002-9711-9424]{Ashley Bransgrove}\thanks{E-mail:  abransgrove@princeton.edu}
\affiliation{Princeton Center for Theoretical Science and Department of Astrophysical Sciences, Princeton University, Princeton, NJ 08544, USA}

\begin{abstract}
Crust quakes are frequently invoked as a mechanism to trigger sudden transients in the magnetospheres of magnetars. 
In this picture, a mechanical failure of the crust excites seismic motions of the magnetar surface that launch force-free waves into the magnetosphere.
We first investigate this problem analytically and then perform three-dimensional numerical simulations. 
Our simulations follow the propagation of high-frequency magneto-elastic waves in the entire crust, and include magnetic coupling to the dipolar magnetosphere and liquid core through simplified radiation boundary conditions.  
We observe seismic waves bouncing between the crust-core interface and the surface with a characteristic frequency $\sim 1$~kHz, which could appear as a modulation of the magnetospheric radiation. 
Both the star quake and its associated magnetospheric wave emission are strongly damped on a timescale $\sim 10 \ \rm ms$ by magnetic coupling to the liquid core. 
Since the seismic waves are significantly damped before they can spread laterally around the crust, magnetospheric wave emission occurs primarily near the initial epicenter of the quake. 
Our simulations suggest that non-axisymmetric quakes will launch a mixture of Alfv\'en and fast magnetosonic waves into the magnetosphere. 
The results will be important for interpreting magnetar bursts and understanding the possible trigger mechanisms of fast radio bursts. 
\end{abstract}

\keywords{Magnetars --- Neutron stars --- Magnetic fields --- Plasma Astrophysics}

\section{Introduction}
\label{sec:intro}
Magnetars are strongly magnetized, slowly rotating neutron stars with external dipole magnetic fields $B_\star\simeq10^{14}-10^{15} \ {\rm G}$ \citep{Woods&Thompson2006,Mereghetti2008}. Magnetars display a range of transient radiative activity in the X-ray and soft gamma-ray bands \citep{Harding&Lai2006,Kaspi&Beloborodov2017}. 
The most energetic events are giant flares, which release up to $\sim 10^{46} \ \rm ergs$ in gamma rays in $\sim 0.5$~s \citep{Hurley_2005,Palmer_2005}. Quasi-periodic oscillations (QPOs) in giant flare light curves have been interpreted as seismic oscillations of magnetar interiors \citep{Kaspi&Beloborodov2017}. 
The most frequently occurring magnetar activity is X-ray bursts, which are short bright bursts of hard X-rays with typical duration $10-100 \ \rm ms$. 

Magnetars have also been proposed to explain fast radio bursts (FRBs); 
bright millisecond duration radio bursts which have been a subject of intense observational and theoretical investigations in recent years. 
The breakthrough detection of a Galactic fast radio burst (FRB 200428) \citep{Bochenek2020,CHIME/FRB2020} in association with a hard X-ray burst \citep{Mereghetti20,CKLi21,konus,AGILE} from the Galactic magnetar SGR 1935+2154 suggests that at least some FRBs are produced by magnetars. 
This event was interpreted as a star quake, which launched a series of Alfv\'en wave pulses into the closed field line region of the magnetosphere, producing simultaneous X-ray and radio bursts \citep{Yuan2020}.

Many proposed FRB emission mechanisms, 
both inside and outside the magnetosphere \citep{Zhang2020,Lyubarsky2021,ZhangRMP}, rely on a star quake first launching force-free waves from the magnetar surface.
One can classify them into two force-free waves induced ways:
(1) Alfv\'en wave induced $E_\parallel$ models that apply curvature radiation in the open field line region \citep{Kumar2017,Kumar&Bosnjak2020,Lu2020,Kumar2022}. In the closed field line region, the free electron laser mechanism may also occur by considering the Alfv\'en waves scattering process \citep{Lyutikov2021,Lyutikov&Freund2025}. 
Additionally, Alfv\'en waves can enter the non-linear regime and escape from the magnetosphere to form an ultra-relativistic ejecta \citep{Yuan2022}, which might be responsible for generating the synchrotron maser process \citep{Lyubarsky2014,Beloborodov2017,Beloborodov2020,Metzger2019,Plotnikov&Sironi2019,Sironi2021}.
(2) Fast magnetosonic waves-induced models that apply inverse Compton scattering (ICS) in the open field line region \citep{Zhang2022,Qu&Zhang2023,Qu&Zhang2024}, magnetic reconnection in the compressed current sheet beyond the light cylinder \citep{Lyubarsky2021,Mahlmann2022} 
and synchrotron maser instability at the monster shock front \citep{Chen_2022, Beloborodov_2023, Vanthieghem&Levinson2024}.
Therefore, force-free waves generated by magnetar crust quakes are closely connected to the trigger mechanisms of FRBs.

The propagation of seismic waves in neutron star crusts was first studied in detail by \cite{Blaes1989}. The enormous hydrostatic pressure of the crust implies that the seismic waves are effectively incompressible. On the other hand, the magnetosphere supports both compressible (fast magnetosonic) and incompressible (Alfv\'en) modes. Coupling of seismic waves to magnetospheric oscillations was studied analytically \citep{Blaes1989}, and later numerically \citep{Bransgrove2020,Yuan2020,Yuan2021,Yuan2022}\footnote{The response of the magnetosphere to internal dynamics was also studied in the slow `quasi-static' limit by \citep{Gabler_2014, Akgun_2017}.}.
However, these studies were limited to axisymmetry, or considered special polarizations of the seismic waves such that only the Alfv\'en mode was excited. 
In principle, fast modes could also be emitted directly from the neutron star surface, depending on the polarization of the seismic wave and the orientation of the background magnetic field \citep{Beloborodov_2023}.

The emission of magnetospheric waves from a quake is complicated by several effects: 
(i) For low frequency crustal waves $\omega \lesssim c/R_{\star} \sim 2\times 10^4$~rad~s$^{-1}$ the magnetosphere adjusts in a quasi-static manner near the star \citep{thompson_soft_1995,Gabler_2014}. The wave launching then occurs from the effective radius $r_w\sim R_{\star} + c/\omega$ \citep{Beloborodov_2023}, and this process is not understood in detail. 
(ii) The wave vector immediately above the surface is not purely radial, but can contain a horizontal component $k_h$ due to horizontal gradients of the surface velocity (e.g. \cite{Bransgrove2020}). 
(iii) The magnetospheres of active magnetars are probably strongly twisted with significant gradients in the magnetic field structure that can affect wave propagation.
In this work we focus primarily on the three-dimensional dynamics and energetics of seismic waves inside the crust. 
For simplicity we consider a dipolar magnetosphere, and limit ourselves to high-frequency crustal waves $\omega>c/R_{\star}$. We also consider crustal deformations with large-scale angular structure, so that the horizontal wave vector satisfies $k_h \ll \omega/c$, and subsequently $\vec{k}\approx k_r \vec{\hat{r}} $. Wave launching by more general crustal deformations requires further investigation using force-free simulations of the magnetosphere.

In this paper, we investigate the three-dimensional dynamics of magneto-elastic waves in magnetar crusts. First, using a simplified one-dimensional (1D) model, we consider analytically the excitation and propagation of waves in the crust. Using the simplified model, we show how both Alfv\'en and fast magnetosonic waves are launched by seismic motions of the magnetar surface. We then extend previous work by deriving the transmission efficiency of seismic disturbances into fast magnetosonic waves in the magnetosphere for the first time. Then using full three-dimensional numerical simulations we explore the magneto-elastic dynamics of a magnetar crust and its coupling to the magnetosphere and core.  
Our model shows that the star quake and magnetospheric wave emission are quickly damped by the emission of Alfv\'en waves into the magnetar core.

This paper is organized as follows. 
In Section~\ref{sec:structure}, we describe the physical properties of our model neutron star and the equations of motion. 
In Section~\ref{sec:waves_from_quakes}, we outline a simplified 1D analytic model of seismic waves and their coupling to the magnetosphere and the core in Cartesian geometry. 
We also calculate the transmission efficiency of seismic disturbances into Alfv\'en and fast magnetosonic waves in a strongly magnetized neutron star.
In Section~\ref{sec:setup}, we describe the setup of our numerical simulations, and in Section~\ref{sec:numerical results}, we show global three-dimensional numerical simulations that illustrate the star quake dynamics and wave emission in realistic geometry. 
The conclusions and observational implications are discussed in Section~\ref{sec:Conclusion}.
Throughout the paper, the convention $Q=10^n Q_n$ is adopted in cgs units.

\section{Equations of Motion of the Crust}\label{sec:structure}

\subsection{Magneto-elastic Wave Equation}
The linearized momentum equation and the continuity equation are given by \citep{Blaes1989}
\begin{equation}
\begin{aligned}\label{eq:motion equation}
\rho\ddot{\vec\xi}&=\nabla\cdot\vec\sigma+\frac{1}{c} \vec j\times\vec B+\vec g \delta\rho-\nabla \delta p,
\end{aligned}
\end{equation}
and
\begin{equation}
\delta\rho +\nabla\cdot(\rho\vec \xi)=0, 
\end{equation}
where $\vec{\xi}$ is the lagrangian displacement of an element of material, $\vec g$ is the gravitational acceleration, $\delta p$ and $\delta \rho$ are the perturbed pressure and density, and $\vec\sigma$ is the elastic stress tensor of an isotropic incompressible solid 
\citep{Landau1959}
\begin{equation}\label{eq:stress tensor} \sigma_{ij}=\frac{\mu}{2}\left(\frac{\partial\xi_i}{\partial x_j}+\frac{\partial \xi_j}{\partial x_i}\right),
\end{equation}
where we have neglected terms of second order. Throughout this work we assume small deformations such that the strain $\epsilon\ll 1$, and linear elasticity theory is valid. The plastic response of the crust at large strains is beyond the scope of this work. 
We assume that radial motions are suppressed by stratification ($\xi_r = 0$), and that the allowed perturbations are incompressible because the hydrostatic pressure far exceeds the elastic shear modulus ($\nabla\cdot\vec{\xi} = 0$). 
This implies $\delta\rho = 0$ and $\delta p = 0$. 
On the timescales of interest, the crust can be considered a perfect conductor. 
The induced electric field is then
$\vec {\delta E}=-\dot{\vec \xi}\times\vec B/c$.
By using the Faraday equation $\partial \vec {B}/\partial t=-c\nabla\times\vec E$, one can obtain
\begin{equation}\label{eq:B_w equation}
\vec {\delta B}=\nabla\times(\vec \xi\times\vec B),
\end{equation}
which describes the perturbation of the background magnetic field and its relation to the crustal displacement $\vec{\xi}$.
The magneto-elastic dynamic wave equation inside the crust is then given by \citep{Blaes1989,Bransgrove2020}
\begin{equation}\label{eq:elasto-dynamic wave equation}
\begin{aligned}
\rho\ddot{\vec \xi}+\rho_B\ddot{\vec \xi}_\perp=&(\nabla\mu\cdot\nabla)\vec\xi-(\vec\xi\cdot\nabla)\nabla\mu+\mu\nabla^2\vec\xi\\
&+\frac{1}{4\pi}\{\nabla\times[\nabla\times(\vec\xi\times\vec B)]\}\times\vec B,
\end{aligned}
\end{equation}
where $\rho_B=B^2/4\pi c^2$ denotes the effective mass density of the background magnetic field and $\ddot{\vec \xi}_\perp=\ddot{\vec\xi}-(\ddot{\vec\xi}\cdot\vec{\hat{B}})\vec{\hat{B}}$ is the acceleration due to the displacement perpendicular to the background magnetic field $\vec B$.
The first three terms on the right side of Equation~(\ref{eq:elasto-dynamic wave equation}) describe elastic restoring force and the last term describes the magnetic restoring force (Ampere force).

\subsection{Magnetar Interior Model}
In this section, we describe the magnetar inner structure used in our calculations. We closely follow the setup of \citep{Bransgrove2020}. We consider a magnetar with dipole magnetic field strength $B_{\rm dip}=4\times 10^{14}$ G. The effective mass density of the magnetic field near the surface is
\begin{equation}
\rho_B=\frac{B^2}{4\pi c^2}\approx  (10^7 \ {\rm g \ cm^{-3}}) \ b^2,
\end{equation}
where $b = B / (4\times 10^{{14}} \ {\rm G})$. 
This defines the transition from the outer liquid envelope of the neutron star to its magnetosphere. 
The top of the crust is defined by the solid–liquid boundary, which is controlled by the Coulomb coupling parameter 
\begin{equation}
\Gamma=\frac{Z^2e^2}{ak_BT},
\end{equation}
where $a=(4\pi n_i/3)^{-1/3}$ is the mean separation distance of the ion lattice, 
$n_i$ is the number density of ion in the crust, 
$T$ is the temperature, and $Z$ is the atomic number.
The liquid layer freezes into a  crystal at $\Gamma\simeq175$ \citep{Potekhin&Chabrier2000}. 
This sets the crystallization density $\rho_{\rm crys}$ as a function of temperature 
\begin{equation}
\begin{aligned}
\rho_{\rm crys} \approx   
(10^{11} \ {\rm g \ cm^{-3}}) \ T_9^3\left(\frac{Z}{26}\right)^{-6}\left(\frac{A}{56}\right),
\end{aligned}
\end{equation}
where $A$ as the ion mass, and we adopt $T=10^9$ K as the typical internal temperature of a magnetar.

The hydrostatic density profile of the neutron star is obtained by integrating the TOV equations with the SLy equation of state \citep{douchin_unified_2001} with a central density of $\rho = 10^{15}$~g~cm$^{-3}$. We use the analytical fitting formula of \citep{haensel_analytical_2004}. 
This gives a neutron star of mass $M_\star=1.4 M_\odot$ and radius $11.69$~km. 
The crust-core boundary is located at $r_{\rm core} = 10.8$~km and density $\rho_{\rm core} = 1.27\times 10^{14} $~g~cm$^{-3}$.

\begin{figure}
\includegraphics[width=90mm]{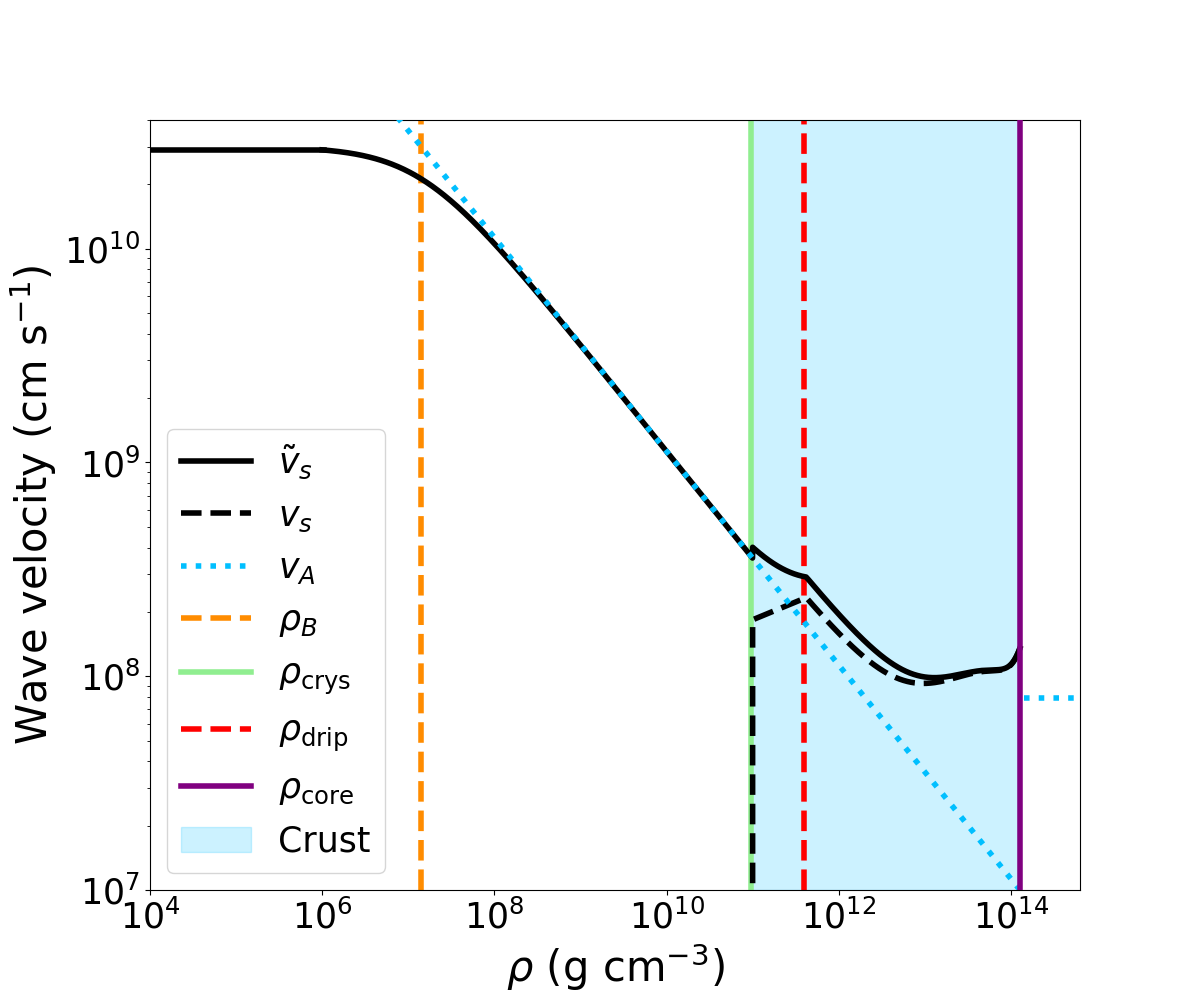}
\caption{The values of effective wave speed ($\tilde v_s$, black line) and Alfv\'en wave speed ($v_A$, blue dashed line) as a function of mass density ($\rho$) in the magnetar magnetosphere, ocean, crust and core. 
Characteristic mass densities are presented.
The following parameters are adopted: magnetic field strength $B_\star=4\times10^{14} \ \rm G$, a magnetar of mass $M_\star = 1.4$~$M_\odot$, magnetar radius $11.69 \ \rm km$.
The crust-core boundary is located at $r_{\rm core} = 10.8 \ \rm km$, and the mass density at the core $\rho_{\rm core}=1.27\times10^{14} \ {\rm g \ cm^{-3}}$.
}
\label{fig:wave_speed}
\end{figure}

The shear modulus of the crust is proportional to the Coulomb energy of the lattice and is given by \citep{Ruderman1968,Strohmayer1991}
\begin{equation}
\mu=0.295Z^2e^2n_i^{4/3}. 
\end{equation}
In the outer crust $\rho<\rho_{\rm drip} = 4\times 10^{11}$~g cm$^{-3}$, the shear modulus scales as $\mu\propto \rho^{4/3}$. In the inner crust, 
with mass density $\rho_{\rm drip}<\rho <\rho_{\rm core}$, 
the modulus approximately scales as $\mu\propto \rho$. 
Thus, the elastic wave speed $ v_s \equiv (\mu/\rho)^{1/2}\sim 10^8 \ \rm cm \ s^{-1}$ is effectively constant throughout the crust (Figure~\ref{fig:wave_speed}), 
and the elastic wave crossing time is given by $\tau_c \sim H/v_s \sim 1$ ms, where $H\sim0.1R_{\star}\sim 10^5 \ \rm cm$ is the thickness of the crust and $v_s \sim 10^{8} \ \rm cm \ s^{-1}$ is the characteristic elastic wave speed. 
The characteristic densities $\rho_B$, $\rho_{\rm crys}$, $\rho_{\rm drip}$, and $\rho_{\rm core}$ are shown in Figure~\ref{fig:wave_speed}.

Relativistic degenerate electrons dominate the pressure at densities $\rho\lesssim 6\times10^{12} \ {\rm g \ cm^{-3}}$.
The pressure of the electron gas is $P_e\simeq(1.2\times10^{31} \ {\rm erg \ cm^{-3}}) \ \rho_{12}^{4/3}Y_e^{4/3}$, where $Y_e$ denotes the number fraction of electrons.
For analytic calculations, we assume the electron number fraction $Y_e=0.35$ in the outer crust where $\rho\lesssim\rho_{\rm drip}$.    
The hydrostatic equilibrium then gives the relation between the mass density and depth as $|z|\approx 5\times 10^{4} \rho_{12}^{1/3}$~cm.

\section{1D Model of Elastic Waves}\label{sec:waves_from_quakes}

In this section, we discuss the generation of both Alfv\'en and fast waves in the simplest possible configuration of a one-dimensional (1D) plane-parallel Cartesian crust. 
In Section~\ref{sec:elastic_waves} and \ref{propagation}, we discuss the excitation of shear waves by a star quake. 
In Section~\ref{sec:coupling_to_magnetosphere} we describe the coupling to magnetospheric waves, and in Section~\ref{sec:transmission_mag}, we calculate the transmission of elastic waves through the surface layers of the crust.
In Section~\ref{sec:transmission_core}, we calculate the transmission of elastic waves into the magnetar core and the lifetime of the star quake.

\begin{figure}[t!]
\includegraphics[width=86mm]{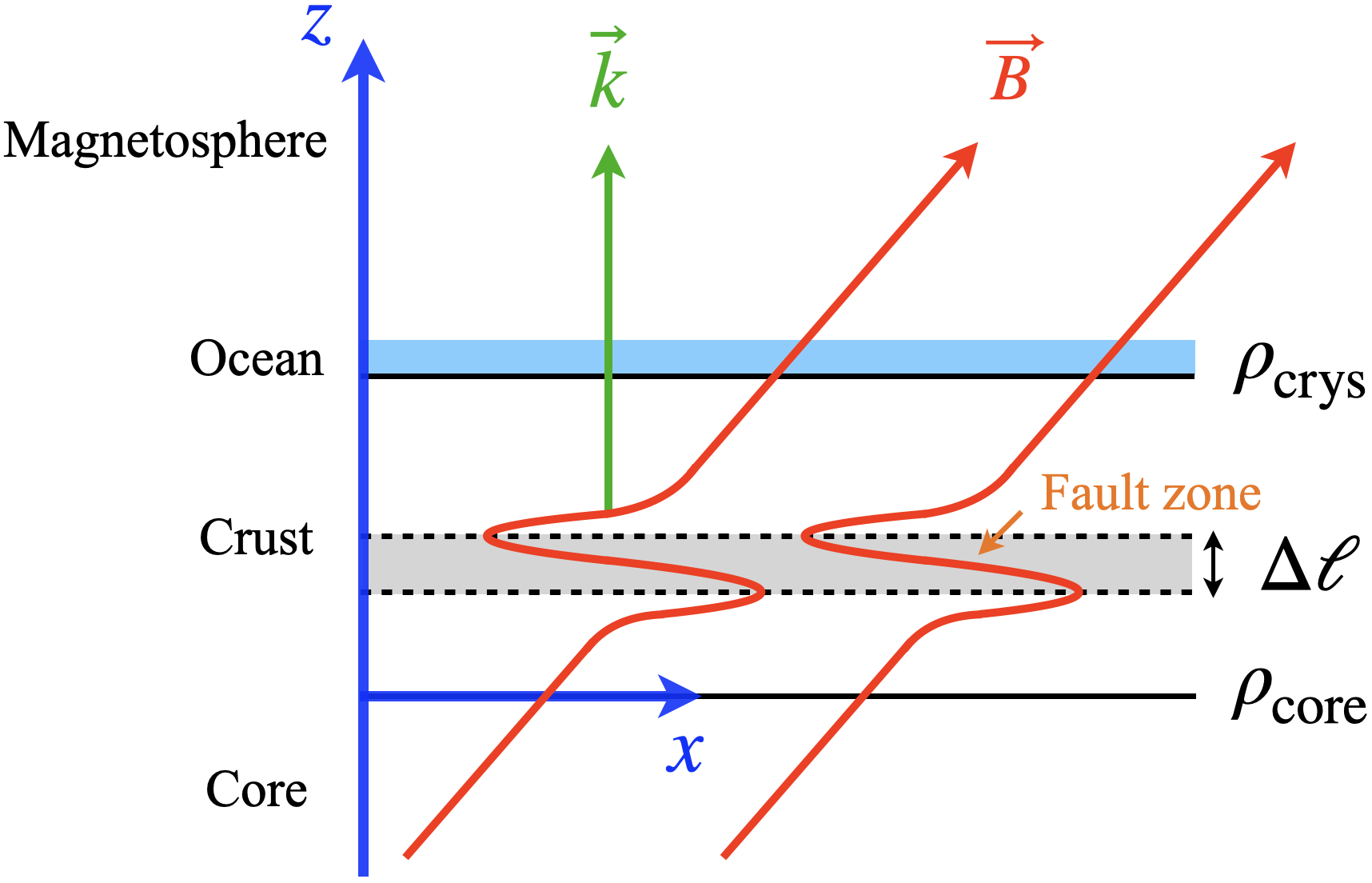}
\caption{Toy model of a crust quake. The incompressible displacement of field lines $\xi_x$ (red wiggler) is generated by the sudden yielding of a strain layer in the deep crust,
and propagates vertically along $z$-axis with wave vector $k_z$ (green arrow) towards the magnetar surface.
The blue shaded region shows the liquid ocean and the gray shaded region shows the fault zone of thickness $\Delta \ell$. 
}
\label{fig:cartoon_pic}
\end{figure}

The main components of our simplified 1D model are described in Figure~\ref{fig:cartoon_pic}. 
We assume that the wave propagates vertically ($\vec{k}=k_z \vec{\hat{z}}$), and all physical quantities depend only on the $z$-coordinate ($\partial_x = \partial_y = 0$). 
Inside the magnetar crust, we consider a uniform time independent background magnetic field that lies in the $x-z$ plane with angle $\theta_0$ with respect to the $z$-axis
\begin{equation}
\vec B=B_x\vec{\hat{x}}+ B_z\vec{\hat{z}},
\end{equation}
where $B_x = B\sin\theta_0$, $B_z = B\cos\theta_0$. 
Strong radial stratification suppresses vertical displacements  ($\xi_z = 0$),
and the allowed the horizontal displacements are incompressible ($\nabla\cdot \vec{\xi} =0$) due to the enormous hydrostatic pressure.
Therefore, we assume a displacement of the form
\begin{equation}
\vec{\xi}(z,t) = \xi_x (z,t)\vec{\hat{x}} +  \xi_y (z,t)\vec{\hat{y}}.
\label{xi_1D}
\end{equation}
Note that $\vec{\xi}(z,t)$ automatically satisfies the incompressibility condition in this 1D problem since $\partial \xi_x/\partial x = \partial \xi_y/\partial y = 0$.

\subsection{Generation of Shear Waves by a Star Quake}
\label{sec:elastic_waves}

Following \cite{Bransgrove2020}, we model the quake as the sudden yielding of a strain layer in the deep crust. 
The thickness of the strain layer is $\Delta\ell $, a fraction of the local pressure scale height $\sim 3\times 10^4 \ \rm cm$. 
The sudden failure of the strain layer of thickness $\Delta \ell$ excites elastic waves with characteristic angular frequency $\omega_Q \approx v_s / \ell_0$, where $v_s\simeq \tilde v_s$ in the deep crust (Figure~\ref{fig:wave_speed}) and $\ell_0\sim \Delta \ell $ is the characteristic scale of the deformation gradient \citep{Bransgrove2020}. 
This gives
\begin{equation}
\omega_Q \approx \frac{v_s}{\ell_0}\sim (10^5~\text{rad~s}^{-1})\left( \frac{\Delta \ell}{3\times 10^3~\text{cm}}\right)^{-1},
\label{frequency}
\end{equation}
where we have taken $v_s \approx 2\times 10^8$~cm~s$^{-1}$ at $\rho \gtrsim \rho_{\rm drip}$ (see Figure~\ref{fig:wave_speed}). 
The elastic waves cross the thickness of the crust in $\tau_c \sim H/v_s \sim 1 \ \rm ms$, and experience several reflections at the  surface and the crust-core due to the impedance mismatch \citep{Blaes1989}. 
The characteristic frequency of this `bouncing' is $\nu_b\sim v_s/H\sim 1 \ \rm kHz$. 
In a multidimensional setting, the waves also spread sideways around the crust \citep{Bransgrove2020}. 
The energy released by the quake is 
\begin{equation}\label{eq:E_Q}
E_Q = \frac{1}{2}\mu\epsilon_{\rm yield}^2\Delta \ell A \approx (10^{42}  \ {\rm erg}) \ A_{Q,11}\left(\frac{\Delta \ell}{3\times10^{3} \ \rm cm}\right),
\end{equation}
where $\mu\approx 10^{30} \ \rm erg \ cm^{-3}$ is the shear modulus of the deep crust, $\epsilon_{\rm yield}\sim 0.1$ is the breaking strain\footnote{If the deep crust contains a layer of nuclear pasta the breaking strain could be even larger \citep{Caplan_2018}.} \citep{Horowitz&Kadau2009}, and $A_Q$ is the horizontal area of the fault plane.

\subsection{Propagation of Waves in the Crust}
\label{propagation}

\begin{figure*}
\includegraphics[width=18 cm,height=6 cm]{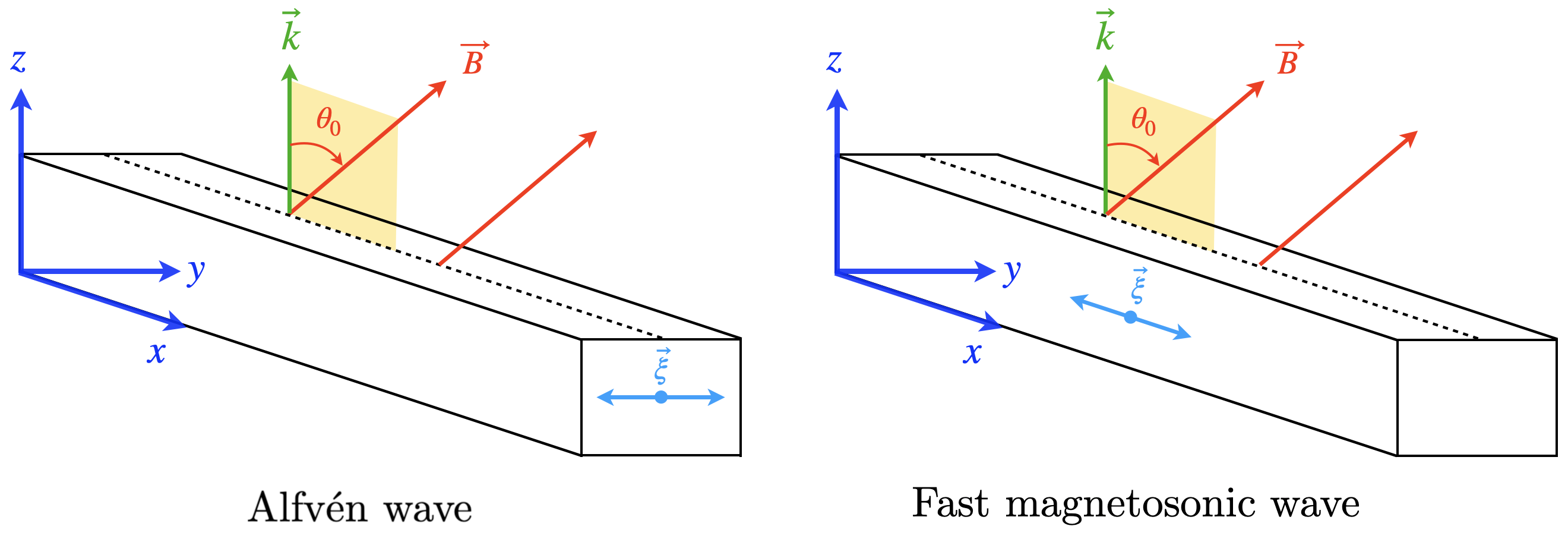}
    \caption{Relative directions of the background magnetic field $\vec B$ (red arrow), wave vector $\vec k$ (green arrow), and displacement perturbation of elastic waves $\vec \xi$ (blue arrow) in the crust of the magnetar.  
    The wave vector $\vec k$ is aligned along the $z$-axis.
    $\vec B$ is assumed to be in the $x-z$ plane (indicated by the yellow plane) with the angle $\theta_0$ with respect to the $z$-axis.
    The displacement along the $y$-axis generates pure Alfv\'en waves (left panel), whereas displacement along the $x$-axis generates pure fast magnetosonic waves (right panel).}
\label{fig:coordinate}
\end{figure*}

In the 1D model the magneto-elastic wave Equation~(\ref{eq:elasto-dynamic wave equation}) is significantly simplified.
The $x$-component of displacement $\xi_x(z,t)$ is parallel to the horizontal component of $\vec{B}$ and evolves according to
\begin{equation}\label{eq:wave equation xi along x}
\tilde{\rho}\frac{\partial^2\xi_x}{\partial t^2}=\frac{\partial}{\partial z}\left(\tilde\mu\frac{\partial\xi_x}{\partial z}\right),
\end{equation}
where $\tilde\rho$ and $\tilde\mu$ are the effective mass density and the effective shear modulus, respectively, which are given by
\begin{equation}
\tilde\rho=\rho+\rho_{Bz}=\rho+\frac{B_z^2}{4\pi c^2}, \ \tilde\mu=\mu+\frac{B_z^2}{4\pi}.
\end{equation}
The $y$-component of displacement $\xi_y(z,t)$ is completely perpendicular to $\vec B$ and evolves according to
\begin{equation}\label{eq:wave equation xi along y}
\tilde{\rho}\frac{\partial^2\xi_y}{\partial t^2}=\frac{\partial}{\partial z}\left(\tilde\mu\frac{\partial\xi_y}{\partial z}\right),
\end{equation}
where 
\begin{equation}
\tilde\rho=\rho+\rho_B=\rho+\frac{B^2}{4\pi c^2}, \ \tilde\mu=\mu+\frac{B_z^2}{4\pi}.
\end{equation}
One can see that Equations~(\ref{eq:wave equation xi along x}) and (\ref{eq:wave equation xi along y}) share a similar form, 
with the only distinction that $\tilde\rho$ depends on $B_z$ only in Equation~(\ref{eq:wave equation xi along x}),
whereas $\tilde\rho$ depends on $B$ in Equation~(\ref{eq:wave equation xi along y}).
The magneto-elastic wave speeds of $\xi_x$ and $\xi_y$ are
\begin{equation}\label{eq:v_s_x}   
\tilde v_{s,x}=\left(\frac{\tilde{\mu}}{\tilde{\rho}}\right)^{1/2} =\left(\frac{\mu+{B_z^2}/{4\pi}}{\rho+{B_z^2}/{4\pi c^2}}\right)^{1/2},
\end{equation}
and
\begin{equation}\label{eq:v_s_y}   
\tilde v_{s,y}=\left(\frac{\tilde{\mu}} {\tilde{\rho}}\right)^{1/2}=\left(\frac{\mu+{B_z^2}/{4\pi}}{\rho+{B^2}/{4\pi c^2}}\right)^{1/2}.
\end{equation}
In the magnetosphere, we have $\rho=0$ and $\mu=0$. 
The wave speeds for perturbations in the $x$ and $y$ directions are then 
\begin{equation}
\tilde v_{s,x}=c, \ \tilde v_{s,y}= c \cos\theta_0,
\end{equation}
corresponding to the vertical group velocity of the fast magnetosonic wave (fast mode) and the Alfv\'en wave -- the two eigenmodes of force-free electrodynamics [e.g. \citep{komissarov_electrodynamics_2004}]. 
The electric field of the Alfv\'en wave is polarized in the $\vec k-\vec B$ plane ($x-z$ plane) and propagates along $\vec B$. 
The fast mode is polarized perpendicularly to the $\vec k-\vec B$ plane ($y$-direction), and propagates freely along $\vec k$.
The dispersion relations of the two waves are
\begin{equation}
\omega^2=k^2c^2
\end{equation}
for fast modes and
\begin{equation}
\omega^2=k^2c^2\cos^2\theta_0
\end{equation}
for Alfv\'en waves. 
This suggests that displacements of the crust in the $x$-direction couple to the fast mode, and displacements of the crust in the $y$ direction couple to the Alfv\'en wave. 
We will explore this coupling further in the following section.

Note that in the steady state when the waves propagate in the WKB regime, the energy flux of the magneto-elastic waves $F\sim\rho \dot{\xi}^2 \tilde v_s\sim \rho \omega^2\xi^2 \tilde v_s =$~const. 
This implies the scaling of the wave amplitude $\xi \propto (\rho\tilde{v}_s)^{-1/2}$. For $\rho\lesssim\rho_{\rm crys}$, we have $\tilde{v}_s \approx v_A$, so the amplitude scales as $\xi \propto \rho^{-1/4}$ and the strain $\epsilon = \partial\xi/\partial z \propto \rho^{1/4}$. 
Therefore, the waves are weakly growing as they propagate toward the surface.

\subsection{Coupling the Crust and the Magnetosphere}
\label{sec:coupling_to_magnetosphere}

In this subsection, we describe the coupling of elastic waves in the crust to force-free waves in the magnetosphere. 
In order to investigate the coupling of magnetar surface motions to different wave modes in the magnetosphere, 
we consider a general horizontal perturbed velocity $\vec{v} = \dot{\vec{\xi}}$ inside the crust of the magnetar as
\begin{equation}
\vec v=v_0\cos\phi e^{i\omega t}\hat{x}+v_0\sin\phi e^{i\omega t}\hat{y},
\end{equation}
where ${\rm Re}(v_0e^{i\omega t})=v_0\cos(\omega t)$ denotes the initial perturbation amplitude, $\omega$ is the angular frequency of the perturbation and $\phi$ is the azimuthal angle between the $\vec v$ and $x$-axis.

On the timescale of a crust quake, the crust and the magnetosphere are regarded perfect conductors.
We define the unit normal vector $\hat{n}$ that is perpendicular to the magnetar surface.
From interface conditions for electromagnetic fields, we find
\begin{equation}
\hat{n}\times(\vec E^{\rm out}-\vec E^{\rm in})=0 \ \Rightarrow \ \vec E_{t}^{\rm out}=\vec E_{t}^{\rm in},
\end{equation}
and
\begin{equation}
\hat{n}\cdot(\vec B^{\rm out}-\vec B^{\rm in})=0 \ \Rightarrow \ \vec B_{n}^{\rm out}=\vec B_{n}^{\rm in},
\end{equation}
where $\vec E_t$ is the tangential electric field and $\vec B_n$ is the normal magnetic field. Here, ``in" and ``out" refer to the location immediately below and above the magnetar surface. 
The electric field immediately below the surface can be expressed as
\begin{equation}
\vec{E}^{{\rm in}} = -\frac{1}{c}\vec{v}^{\rm in}\times \vec{B},
\end{equation}
where $\vec E^{\rm in}$ and $\vec{v}^{\rm in}$ are the induced electric field and velocity immediately below the magnetar surface.
The tangential electric field is continuous across the interface. 
Therefore, the $x$ and $y$ components of electric fields below and above the interface can be written as 
\begin{equation}
E_x^{{\rm in}}=-\frac{1}{c} v_y^{\rm in} B_z = E_x^{{\rm out}},
\label{Ex}
\end{equation}
and
\begin{equation}
E_y^{{\rm in}}=\frac{1}{c} v_x^{\rm in} B_z =  E_y^{{\rm out}}.
\label{Ey}
\end{equation}
The continuity of $B_z$ implies that the velocity of field lines is continuous across the interface, 
i.e. $v_x^{{\rm in}} = v_x^{{\rm out}}$ and $v_y^{{\rm in}} = v_y^{{\rm out}}$. By the continuity of $v_y$ and $B_x$ this implies 
\begin{equation}
E_z^\text{in} = \frac{1}{c}v_y^\text{in} B_x = E_z^\text{out}. 
\label{Ez}
\end{equation}
As we have seen in Section~\ref{propagation}, the Alfv\'en wave has electric field in the $\vec{k}-\vec{B}$ plane ($x-z$ plane), and the fast mode has electric field in the $\vec{k}\times\vec{B}$ direction ($y$ direction). 
Therefore, Equations~(\ref{Ex}) and (\ref{Ez}) for $E_x^\text{out}$ and $E_z^{\rm out}$ show that the Alfv\'en wave couples $v_y^{\rm in}$: motions of the surface perpendicular to the background magnetic field. 
Equation~(\ref{Ey}) for $E_y^{\rm out}$ shows that the fast mode is coupled to $v_x^{\rm in}$: motions of the surface parallel to the horizontal part of the background magnetic field.

\subsection{Transmission of Waves into the Magnetosphere}
\label{sec:transmission_mag}

Magneto-elastic waves propagating through the outer layers of the crust experience reflection when the scale height becomes similar to the wavelength. 
The classic work of \cite{Blaes1989} studied this problem in detail for seismic waves with displacement perpendicular to the background magnetic field. 
In this section, we extend the work of \citep{Blaes1989} by considering a more general polarization of the seismic wave. This enables us to study coupling to fast modes in the magnetosphere, in addition to Alfv\'en waves.

The transmission coefficient (ratio of transmitted to incident energy flux) can be written as 
\begin{equation}
{\cal T} _{\rm mag}=\frac{4Z_{\rm crust}Z_{\rm mag}}{(Z_{\rm crust}+Z_{\rm mag})^2},
\end{equation}
where $Z_{\rm crust}$ is the impedance of the upper crust and $Z_{\rm mag}$ is the impedance of the magnetosphere. 
The impedance is defined as the product of the mass density and the wave speed, $Z = \tilde{\rho}\tilde{v}_s$, where $\tilde{v}_s$ is the effective wave velocity. 
When the magnetic field is inclined, Alfv\'en waves and fast magnetosonic waves will have different values of the velocity (Section~\ref{propagation}), and will therefore have different impedance.
The impedance of the two wave modes in the magnetosphere is given by
\begin{equation}
Z_{\rm mag}^{A}=\rho_B c \cos\theta_0 = \frac{B^2}{4\pi c}\cos\theta_0,
\end{equation}
and
\begin{equation}
Z_{\rm mag}^{F}=\rho_{Bz} c= \frac{B^2}{4\pi c}\cos^2\theta_0,
\end{equation}
where $Z_{\rm mag}^{A}$ is associated with displacements in the $y$-direction, and $Z_{\rm mag}^{F}$ is associated with displacements in the $x$-direction (see Figure~\ref{fig:coordinate} and Equations~(\ref{eq:wave equation xi along x}) \& (\ref{eq:wave equation xi along y})). 
High-frequency magneto-elastic waves propagate in the WKB regime until the characteristic length scale for changes in $\tilde{v}_s$ becomes comparable to the wavelength of the elastic wave. 
Significant reflection then occurs. 
The WKB reflection condition can be written as
\citep{Blaes1989,Li&Beloborodov2015,Bransgrove2020}
\begin{equation}\label{eq:reflection condition}
\lambda=2\pi \frac{\Tilde v_s}{\omega}=\frac{\Tilde v_s}{|d\tilde v_s/dz|},
\end{equation} 
where $\lambda$ denotes the wavelength of the magneto-elastic wave. 
The impedance of the crust is
\begin{equation}
Z_{\rm crust} = \rho_{\rm refl}\tilde{v}_s(\rho_{\rm refl}),
\label{zcrust}
\end{equation}
where $\rho_{\rm refl}$ is the mass density at which the WKB reflection condition (Equation~(\ref{eq:reflection condition})) is satisfied. 
In the crust with $\rho \gg \rho_B$,
the effective wave speed is 
\begin{equation}
\begin{aligned}
\tilde v_s&\approx \left(\frac{\mu}{\rho}+\frac{B^2\cos^2\theta_0}{4\pi\rho}\right)^{1/2}.
\label{vs}
\end{aligned}
\end{equation}
The hydrostatic balance in the outer crust gives the relation $|z|\approx 5\times 10^4 \rho_{12}^{1/3}$~cm (see Section~\ref{sec:structure}). 
Using this relation, together with  Equations~(\ref{eq:reflection condition}) and (\ref{vs}), the reflection density is given by
\begin{equation}
\rho_{\rm refl}\approx (2\times10^{11} {\rm g \ cm^{-3}}) \ b^{6/5}\omega_0^{-6/5} (\cos\theta_0)^{6/5},
\label{refl}
\end{equation}
where $\omega_0=\omega/\omega_Q$ (Equation~(\ref{frequency})). 
One can see that the reflection density decreases for higher frequency elastic waves with $\rho_{\rm refl}\sim B^{6/5}\omega^{-6/5}$, which is consistent with the propagation of waves in pulsar crusts \citep{Bransgrove2020}.
For moderate inclinations of the magnetic field lines (e.g. in the polar regions), the second term in Equation~(\ref{vs}) dominates at $\rho_{\rm refl}$. 
The effective wave speed is then
\begin{equation}
\tilde v_s(\rho_{\rm refl})\approx (3\times10^8 \ {\rm cm \ s^{-1}}) \ b^{2/5}\omega_0^{3/5}(\cos\theta_0)^{2/5}.
\end{equation}
This gives the transmission coefficient for Alfv\'en waves and fast waves in the polar region as 
\begin{equation}
{\cal{T}}^{A}_{\rm mag} \approx 5\% \ b^{2/5}\omega_0^{3/5} (\cos\theta_0)^{-3/5},
\label{TA_polar}
\end{equation}
and
\begin{equation}
{\cal{T}}^{F}_{\rm mag} \approx 5\% \ b^{2/5}\omega_0^{3/5} (\cos\theta_0)^{2/5}. 
\label{TF_polar}
\end{equation}
For strongly inclined magnetic field lines (e.g. in the equatorial regions where $\theta_0 \rightarrow\pi/2$), the first term in Equation~(\ref{vs}) becomes dominant in the outer crust. 
The waves then propagate to the crystallization density, where they experience strong reflection at the solid-liquid boundary. 
The crustal impedance (Equation~(\ref{zcrust})) should then be evaluated at $\rho_{\rm crys}\approx 10^{11}$~g~cm$^{-3}$. 
This gives the transmission coefficient for Alfv\'en waves and fast waves in the equatorial region as
\begin{equation}
{\cal{T}}^{A}_{\rm mag} \approx 10\%  \ b^{2}\cos\theta_0,
\label{TA_eq}
\end{equation}
and
\begin{equation}
{\cal{T}}^{F}_{\rm mag} \approx 10\%  \ b^{2}\cos^2\theta_0,
\label{TF_eq}
\end{equation}
where the suppression of ${\cal T}^{A/F}_{\rm mag}$ as $\theta_0 \rightarrow \pi/2$ is due to the vanishing radial magnetic field $B_r$ on the equator. Note that for a dipole magnetosphere, the field line angle $\theta_0$ is related to the spherical polar coordinate $\theta$ through 
\begin{equation}
\cos\theta_{0}=\frac{2\cos\theta}{({3\cos^2\theta+1})^{1/2}}.
\end{equation}
Equations~(\ref{TA_polar}), (\ref{TF_polar}), (\ref{TA_eq}), and (\ref{TF_eq}) show that the transmission of seismic waves into Alfv\'en and fast modes is essentially the same, except for factors of $\cos\theta_0$. Therefore, at high frequencies $\omega>c/R_{\star}$ we generally expect a similar amount of energy to be emitted into each wave mode. Note that we have considered the transmission of high-frequency waves $\omega_Q\sim 10^5$~rad~s$^{-1}$ that experience strong WKB reflection in the outer layers of the crust. Waves of lower frequency can also experience evanescent reflection that occurs deeper in the crust (cf. \cite{Blaes1989}).

\subsection{Transmission of Waves into the Core}
\label{sec:transmission_core}

Elastic motions of the deep crust launch Alfv\'en waves into the liquid core that drain energy from the quake \citep{levin_qpos_2006}. Following \citep{Bransgrove2020}, the transmission coefficient of vertically propagating shear waves at the crust–core interface is
\begin{equation}
{\cal T}_{\rm core}=\frac{4Z_{\rm crust}Z_{\rm core}}{(Z_{\rm crust}+Z_{\rm core})^2}.
\end{equation}
where $Z_{\rm crust}$ and $Z_{\rm core} $ are the impedances of the crust and the core,
\begin{equation}
Z_{\rm crust} = \rho \tilde{v}_s, \quad  Z_{\rm core} = \rho_p v_A. 
\end{equation}
The impedance of the crust is evaluated immediately above the crust-core interface where $\rho=\rho_{\rm core}$ and $\tilde{v}_s \approx \sqrt{\tilde\mu/\tilde\rho}\approx 10^8$~cm~s$^{-1}$. 
The impedance of the core is evaluated immediately below the crust-core interface. 
Here, neutrons are likely superfluid, and a negligible fraction of their mass is coupled to the proton-electron plasma that participates in the Alfv\'en oscillations \citep{van_hoven_hydromagnetic_2008}. 
Therefore, when calculating $Z_{\rm core}$ we use the density of protons $\rho_p = x_p\rho_{\rm core}$, where $x_p\sim 0.05$ is the proton fraction. 
The Alfv\'en speed in the outer core is given by 
\begin{equation}
v_A=\left(\frac{B H_{c1}}{{4\pi\rho_{p}}}\right)^{1/2} \approx (8\times10^7 \ {\rm cm \ s^{-1}}) \ b^{1/2}, 
\end{equation}
and the Alfv\'en crossing time of the core is $\tau_{\rm A}\sim R_{\star}/v_A\sim 30 \ \rm ms$. 
Here we have assumed that protons are in a type-II superconducting state so that the magnetic flux is confined to superconducting flux tubes. 
The enhanced tension of the tubes increases the usual Alfv\'en speed by a factor $\sqrt{H_{c1}/B}$, where $H_{c1}\approx 10^{15}$~G is the lower critical field of superconductivity \citep{baym_superfluidity_1969,easson_stress_1977}.

For typical magnetar parameters 
$Z_{\rm crust}\gg Z_{\rm core}$.
The transmission coefficient of waves into the core is then  
\begin{equation}
{\cal T}_{\rm core}\simeq \frac{4Z_{\rm core}}{Z_{\rm crust}} \approx 20\% \ b^{1/2}.
\label{T_core}
\end{equation}
Since ${\cal T}_{\rm core}>{\cal T}_{\rm mag}^{A/F}$, we conclude that most of the elastic wave energy is transmitted into the liquid core, rather than into the magnetosphere. 
The characteristic timescale for the quake energy to transfer into the core is
\begin{equation}\label{eq:tau_core}
\tau_{\rm core}=\frac{2\tau_c}{{\cal T}_{\rm core}} \sim 10~{\rm ms}. 
\end{equation}

\section{Numerical Setup}\label{sec:setup}

In this section, we describe the setup of our 3D magneto-elastic dynamics simulations and present the main results.

\subsection{Numerical Method}

The magneto-elastic wave Equation~(\ref{eq:elasto-dynamic wave equation}) is solved in 3D using the spectral method described in \cite{Bransgrove2020}. 
We use torsional eigenmodes of the crust $\vec{\xi}_{nlm} = f_{nl}(r)\vec{\hat{r}} \times \nabla Y_{lm}(\theta,\phi)$ that satisfy $\nabla\cdot \vec{\xi}_{nlm} = 0$. Here $Y_{lm}$ is a spherical harmonic function, and the radial functions $f_{nl}(r)$ are found numerically. We first project our initial condition onto the eigenmodes $\vec{\xi}_{nlm}$, and evolve the spectral coefficients over time. 
Our crustal eigenmodes extend from the crust-core interface at $\rho_{\rm core}=1.27\times 10^{14}$~g~cm$^{-3}$ up to $\rho = 0.1\rho_B $. We therefore capture the acceleration of magneto-elastic waves from $\tilde{v}_s\approx10^{8}$~cm~s$^{-1}$ to $\tilde{v}_s\approx c$. 
Our crustal eigenmodes include magnetic stresses that are important for wave transmission in the crust. When calculating our eigenmodes, we use a spherically symmetric radial magnetic field $B_r = B_\star (r/R_{\star})^{-2}$, with $B_\star = 4\times 10^{14}$~G. This significantly simplifies the calculation of the eigenmodes and only slightly modifies the wave dynamics in the outer crust, because $\mu \gg B^2/4\pi$ everywhere in the crust, except for a thin layer near the surface. As a result, the elastic terms dominate over the magnetic terms in Equation~(\ref{eq:elasto-dynamic wave equation}) in most of the crust volume. The magnetospheric coupling is weakly affected by this approximation because we use a dipolar external field to calculate the magnetospheric feedback on the crust.

Our eigenfunctions are normal modes of a free crust that satisfy zero stress boundary conditions. Therefore, coupling to the magnetosphere and the core is introduced as external forces that act on the top and bottom of the crust. At the crust-core interface we include a radiation force that accounts for the emission of Alfv\'en waves into the liquid core. At the surface we include coupling to both Alfv\'en and fast magnetosonic waves in the magnetosphere, and this is described in the following section. Further details of the numerical method can be found in \citep{Bransgrove2020}.

\subsection{Coupling to the core}
Since the crust and core have very high electrical conductivity, oscillations of the crust deform the magnetic field lines, and launch Alfv\'en waves into the core. Here we describe the back reaction of these waves on the crust, and how it is included in our numerical simulations. 

We follow the approach of \cite{Bransgrove2020}. For simplicity, we assume that the radial magnetic field used in the crust smoothly continues into the core. According to the flux freezing condition, the magnetic perturbation immediately below the crust is 
\begin{equation}
    \vec{\delta B}  = \nabla \times (\vec{\xi}\times\vec{B}) = \frac{1}{r}\partial_r(r B_r\vec{\xi}) \approx \frac{B_r}{v_A}\vec{\dot{\xi}},
    \label{dB}
\end{equation}
where we assumed that there are only inward propagating Alfv\'en waves of the form $\vec{\xi}(t + r/v_A)$, with $v_A$ the Alfv\'en speed\footnote{Since the Alfv\'en crossing time of the core is longer than the quake duration $\tau_A > \tau_{\rm core}$, we have treated the core as a perfect reservoir for Alfv\'en waves. This assumption should be relaxed in future works.}. The presence of $\vec{\delta B}$ implies that the Alfv\'en waves extract a momentum flux from the crust
\begin{equation}
\sigma_{rh} = -\frac{B_r \delta B_h}{4\pi}, 
\label{eq:flux1}
\end{equation}
where $B_r$ is the local radial magnetic field, $\delta B_h$ is the perturbation of the magnetic field immediately above the magnetar surface, and $h=\theta,\phi$ labels the horizontal component of the magnetic perturbation. Following \cite{Bransgrove2020}, the force per unit mass required to extract the momentum flux (Equation~(\ref{eq:flux1})) is $f_h \approx \sigma_{rh}/(\rho \Delta)$, where $\Delta$ is the thickness of the thin layer where the force is applied, and $\rho$ the density at the bottom of the crust. 
In the limit $\Delta\longrightarrow 0$, the force per unit mass of the core on the bottom of the crust is given by
\begin{equation}
\vec{f}_\text{core} = -\frac{B_r \vec{\delta B}}{4\pi\rho}\delta(r-r_{\rm core}). 
\label{eq:force1}
\end{equation}
Substituting Equation~(\ref{dB}) into Equation~(\ref{eq:force1}) gives a radiation type force that acts on the bottom of the crust, and accounts for the momentum flux of outgoing Alfv\'en waves. The force Equation~(\ref{eq:force1}) is projected onto the basis functions in our spectral code at the beginning of each simulation, and used throughout as in \cite{Bransgrove2020}. 

\subsection{Coupling to the magnetosphere}
As discussed in Section~3, seismic motions of the neutron star surface launch waves into the magnetosphere. Here we discuss the force of the outgoing magnetospheric waves on the magnetar surface. 
As in the previous section, we first calculate the Maxwell stress of outgoing waves, and then calculate the force required to extract the momentum flux. This gives a magnetospheric force of the form
\begin{equation}
\vec{f}_{\rm mag} = \frac{B_r \vec{\delta B}}{4\pi\rho_{\rm crust}}\delta(r-R_{\star}),
\label{eq:force}
\end{equation}
where $\rho_{\rm crust}$ the mass density at the outer edge of the crust. Here $\vec{\delta B}$ is the magnetic perturbation immediately above the surface. In contrast to the previous section, we will see that the magnetic perturbation in the magnetosphere $\vec{\delta B}$ contains contributions from both Alfv\'enic and fast magnetosonic waves. 
For simplicity we use a dipole magnetosphere, and consider high-frequency crustal waves such that $\omega \gg c/R_{\star}$. Furthermore, we consider initial crustal perturbations with large-scale angular structure such that the horizontal wave vector $k_h \ll \omega /c$, and subsequently $\vec{k}\approx k_r \vec{e}_r$ immediately above the surface. 
In order to formulate the magnetospheric coupling as a radiation force that acts on the crust, we assume that all emitted waves escape to infinity. 
This assumption is physically motivated by force-free electrodynamics simulations of perturbed magnetospheres \citep{Yuan2020,Yuan2022}. 
These models show that for sufficiently large amplitudes of the crustal perturbations, the Alfv\'en waves become non-linear and ``break out" of the closed magnetosphere, carrying a significant fraction of their energy away from the star, while fast modes easily escape as they propagate freely across the magnetospheric field lines. 
Also note that Alfv\'en waves that return to the neutron star surface will be strongly sheared, and therefore the magnetospheric field lines will not pull coherently on the surface to excite large-scale motions of the crust \citep{levin_qpos_2006,Bransgrove2020}.  
These assumptions significantly simplify the magnetospheric coupling, and allow us to study the dynamics of seismic waves inside the crust while accounting for energy loss through the surface with a radiation force. 
We emphasize that wave emission from more general crustal deformations requires further investigation using dynamical simulations of the magnetosphere.

\begin{figure*}[t!]
\centering
\includegraphics[width=1.0\textwidth]{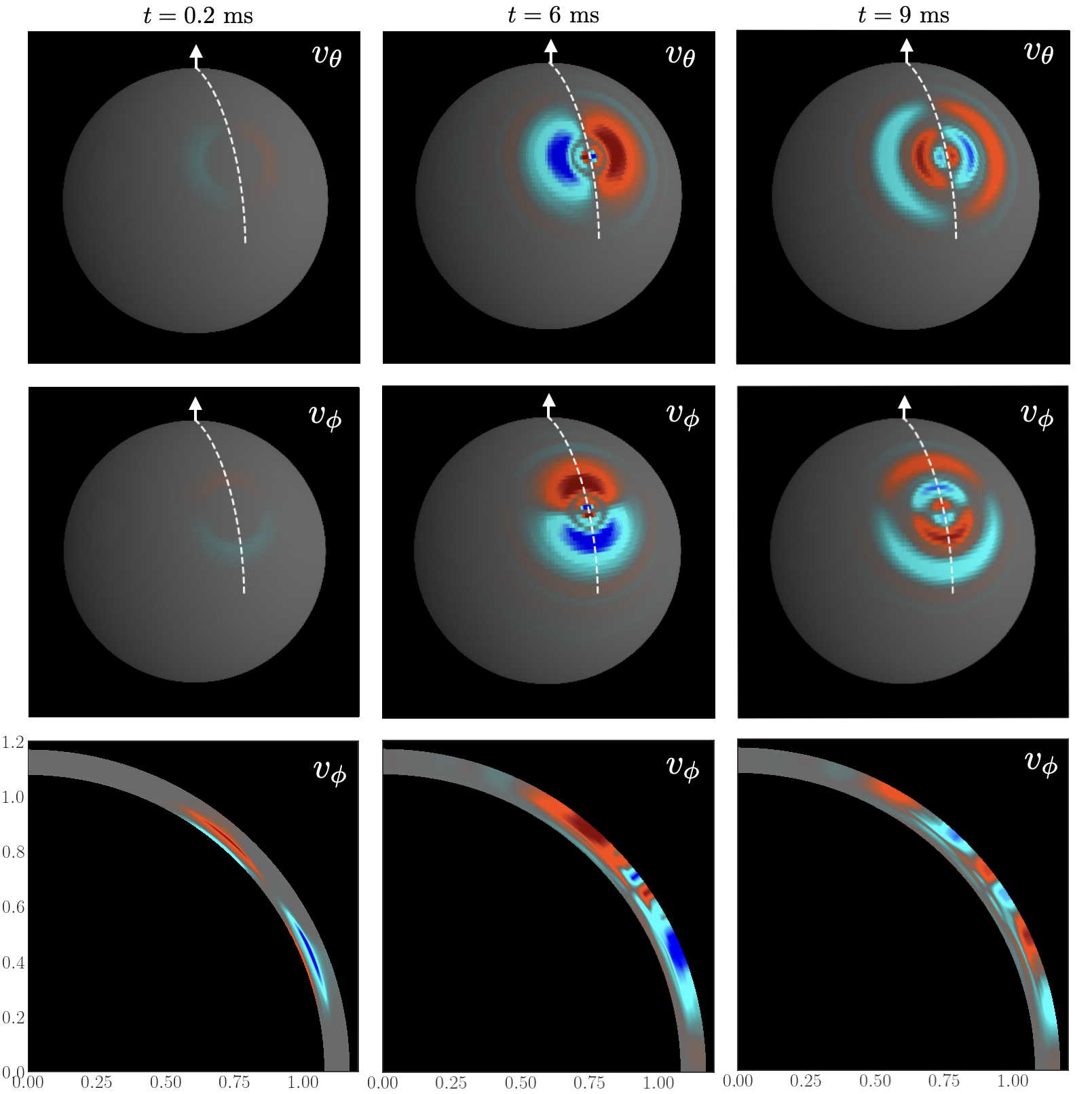} 
    \caption{Evolution of the star quake triggered in the deep crust at $\theta_Q = 3\pi/10$. The left column shows a snapshot during the first elastic crossing time at $t=0.2$~ms, the middle column at $t=6$~ms, and the right column at $t=9$~ms. The first and second row show the velocity fields $v_\theta$ and $v_\phi$ on the neutron star surface. 
    The white arrow indicates the location of the magnetic pole, and the dashed white curve indicates the location of the cross-sectional slice that is displayed in the third row. 
    Red indicates positive velocities, blue indicates negative velocities, and gray indicates zero velocity. 
    All panels are normalized to the same color scale. 
    The axis labels display the $x$ and $z$ coordinate in units of $10 \ \rm km$.}
    \label{quake}
\end{figure*}

We remind the reader that the electric field of the Alfv\'en mode lies in the $\vec{k}-\vec{B}$ plane, and the electric field of the fast mode lies in the $\vec{k}\times\vec{B}$ direction. 
In spherical geometry with a dipole background field, this implies $\vec{E}_{A}= -(\dot{\xi}_\phi B_\theta/c)\vec{\hat{r}} + (\dot{\xi}_\phi B_r/c) \vec{\hat{\theta}}$, and $\vec{E}_{F}  = -(\dot{\xi}_\theta B_r/c) \vec{\hat{\phi}}$ (Alfv\'en waves couple to $\xi_\phi$, and fast modes couple to $\xi_\theta$). 
The corresponding perturbations of the magnetic field are given by the flux freezing condition Equation~(\ref{eq:B_w equation}). The Alfv\'enic perturbation is
\begin{equation}
\vec{\delta B}_{A} \approx B_r \frac{\partial \xi_\phi}{\partial r}\vec{\hat{\phi}} = - \frac{B_r \dot{\xi}_\phi}{c \cos\theta_0 }\vec{\hat{\phi}}
\label{BA}
\end{equation}
where we have assumed an outward propagating Alfv\'en wave of the form $\xi_\phi [t - r/(c \cos\theta_0)]$. The fast magnetosonic perturbation is
\begin{equation}
\vec{\delta B}_{F}  \approx B_r \frac{\partial \xi_\theta}{\partial r}\vec{\hat{\theta}} = - \frac{B_r \dot{\xi}_\theta }{c} \vec{\hat{\theta}}
\label{BX}
\end{equation}
where we have assumed an outward propagating fast mode of the form $\xi_\theta (t - r/c)$. The total magnetic perturbation $\vec{\delta B} = \vec{\delta B}_{A} + \vec{\delta B}_{F}$ then  determines the force on the neutron star surface according to Equation~(\ref{eq:force}). Equation~(\ref{eq:force}) is projected onto the spectral basis functions and included in the evolution equations for the spectral coefficients.
Further details are described in \citep{Bransgrove2020}.

\section{Results}\label{sec:numerical results}

\begin{figure*}[t!]
\begin{center}
\begin{tabular}{lll}
\resizebox{57mm}{!}{\includegraphics[]{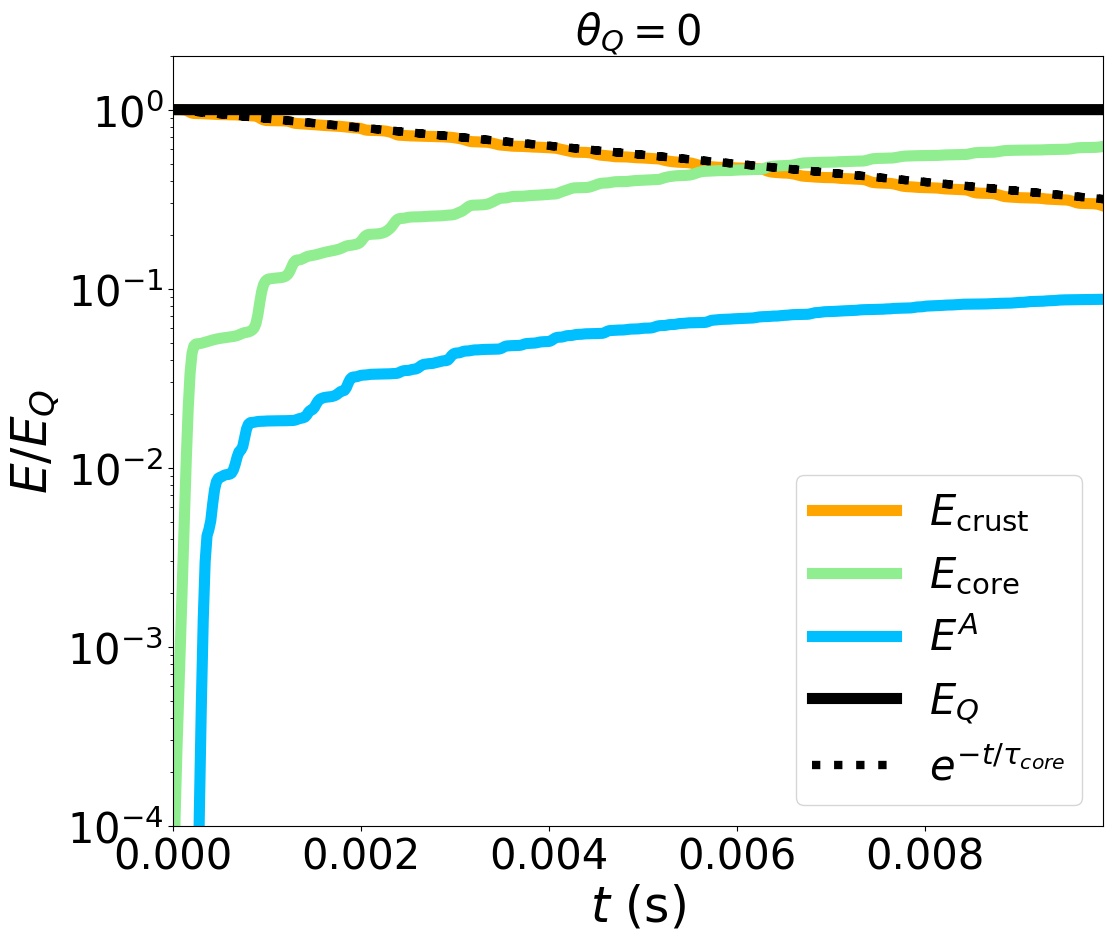}}&
\resizebox{57mm}{!}{\includegraphics[]{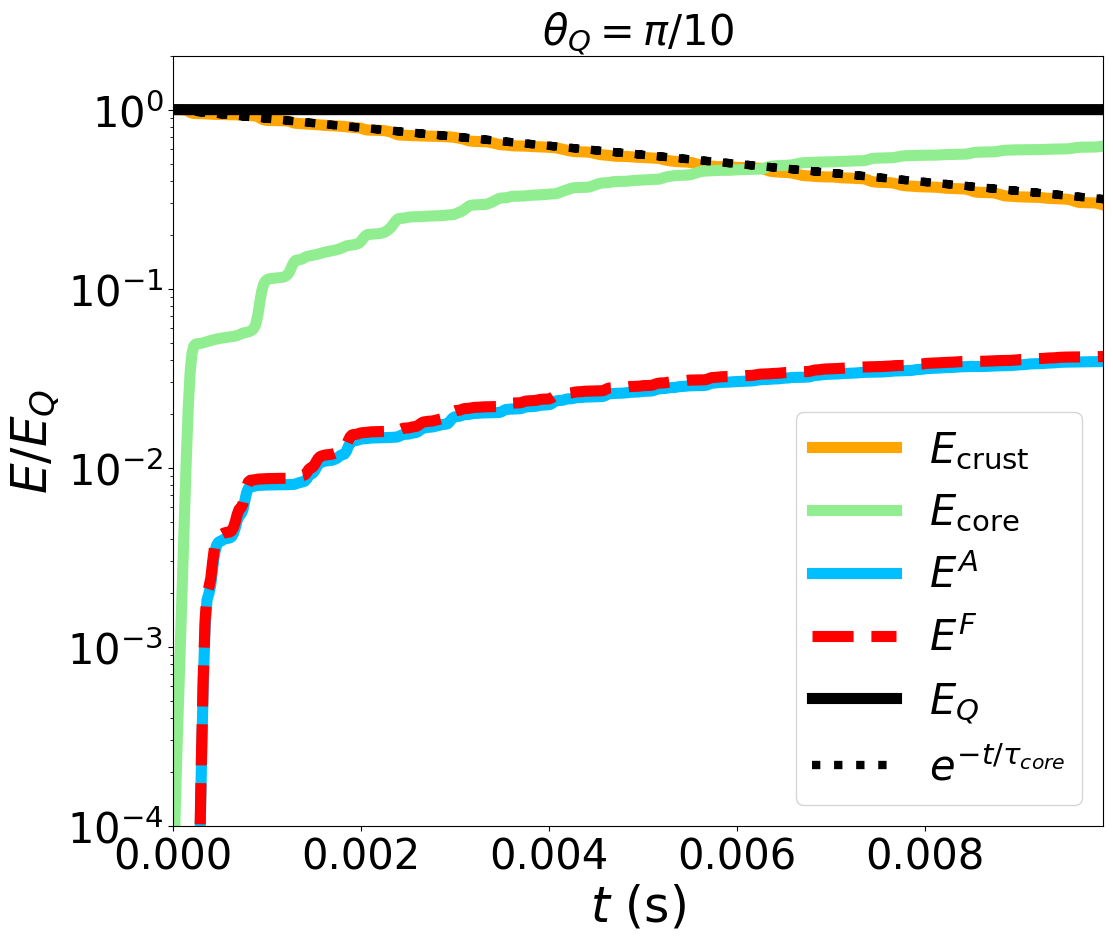}}&
\resizebox{57mm}{!}{\includegraphics[]{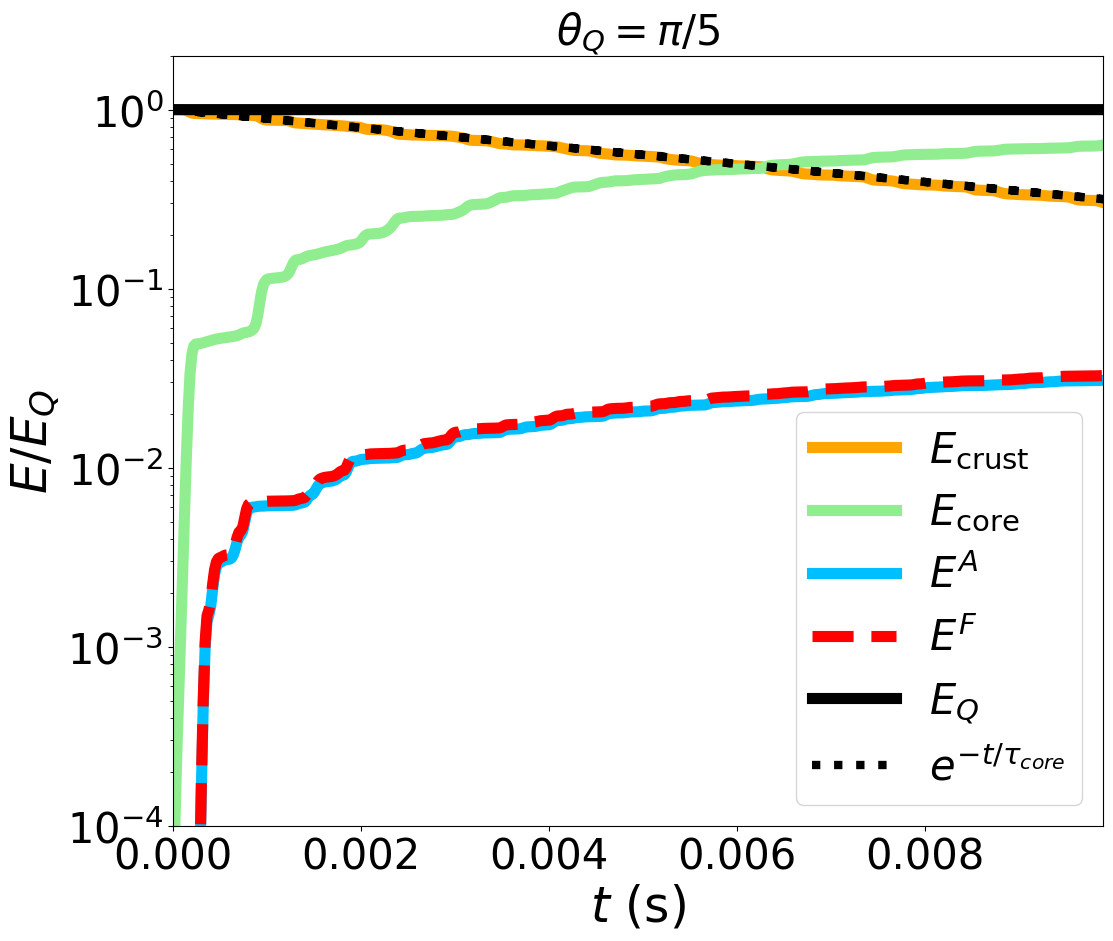}}\\
\resizebox{57mm}{!}{\includegraphics[]{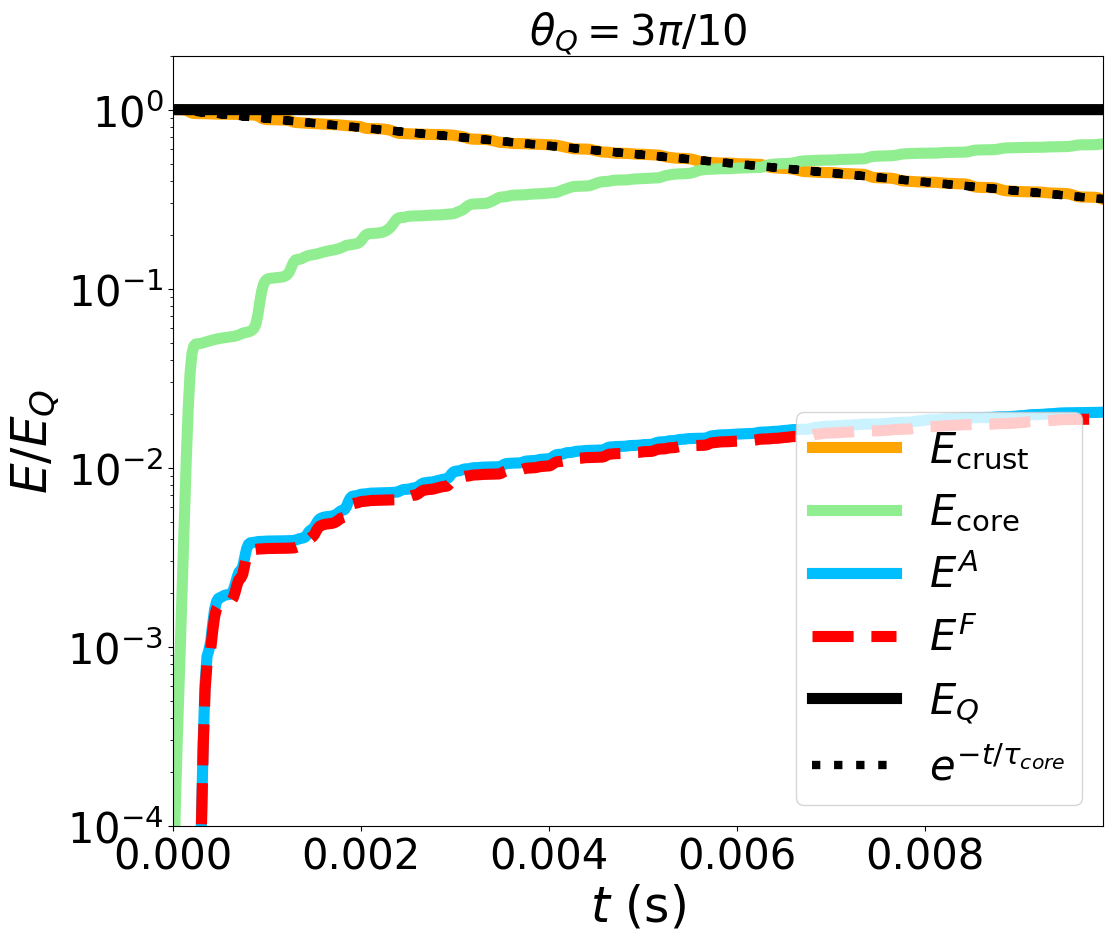}}&
\resizebox{57mm}{!}{\includegraphics[]{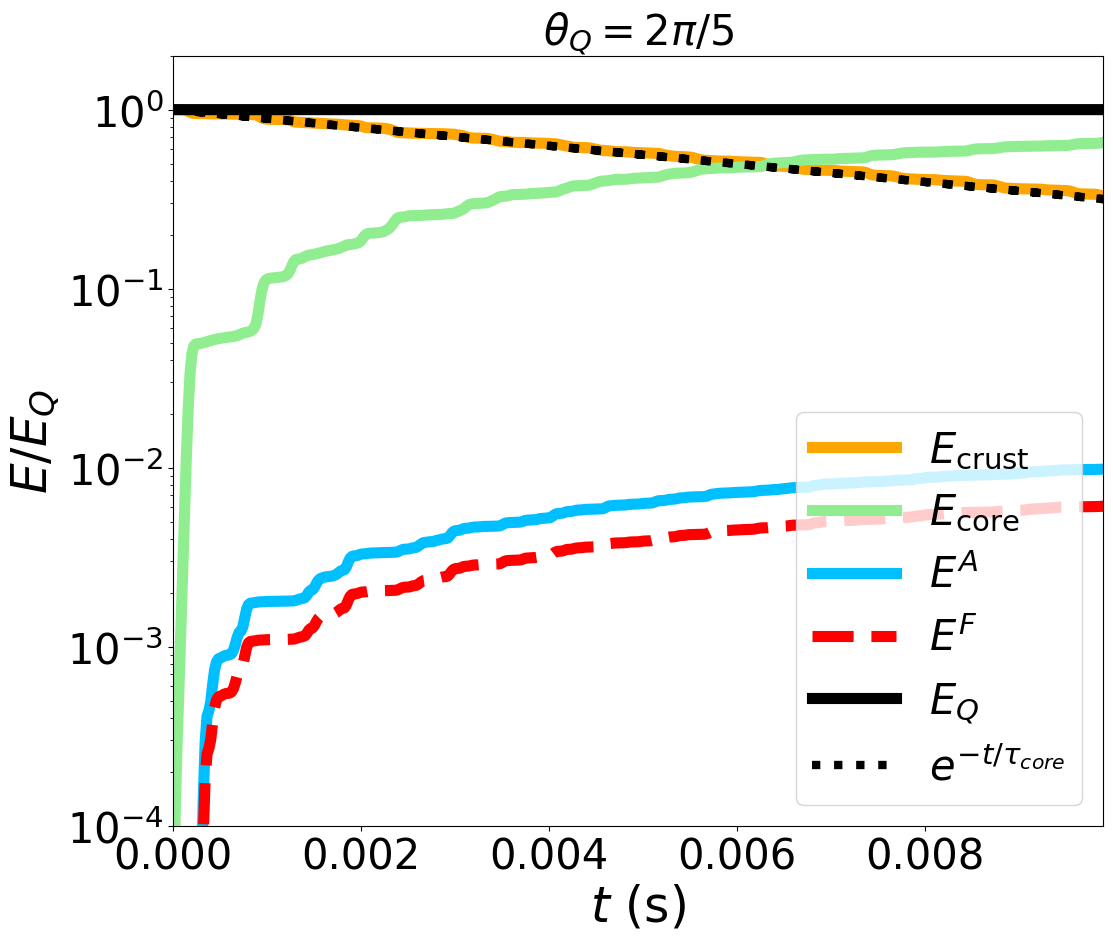}}&
\resizebox{57mm}{!}{\includegraphics[]{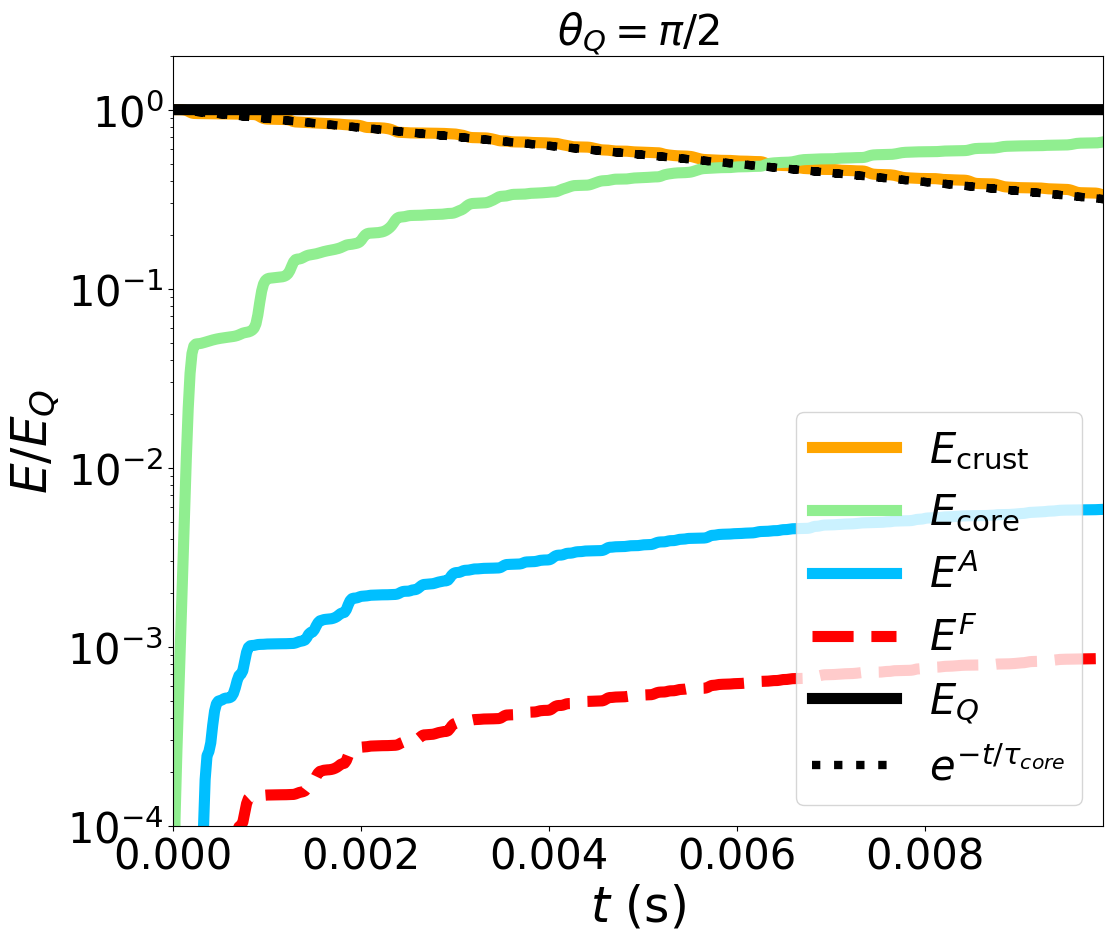}}
\end{tabular}
\caption{Time evolution of energy for different locations of the initial magnetar quake. 
The yellow curve represents the energy in the crust, the green curve represents the energy emitted into the core as Alfv\'en waves, 
and the blue solid and red dashed curves represent the energy emitted into the magnetosphere as Alfv\'en waves and fast magnetosonic waves respectively. 
The dotted black line represents the the analytic model for the decay of quake energy (Equation~(\ref{decay})), and the solid black line demonstrates the conservation of the total quake energy.}
\label{fig:E/E_0alpha} 
\end{center}
\end{figure*}

The initial condition is a smooth twist that is localized in the deep part of the crust, similar to the deformation used by \citep{Bransgrove2020}. 
The half-opening angle of the twisted region is $\Delta\theta\approx \pi / 8$, and the radial thickness of the strain layer is $\Delta \ell\sim 3\times 10^3 \ \rm cm$, which corresponds to the characteristic frequency of the quake $\omega_Q \sim 10^{5}$~rad~s$^{-1}$. 
We first set an axisymmetric twist (pure $\xi_\phi$) that is centered on the polar axis $\theta=0$, then rotate it off the axis to other angles $\theta\neq 0$. 
Rotating the initial quake off the polar axis breaks axisymmetry and generates an initial $\xi_\theta$, 
which allows emission into the fast mode. 
In total we consider 6 locations of the initial of the star quake: $\theta_Q = 0, ~\pi/10,~\pi/5,~3\pi/10,~2\pi/5,~\pi/2$. 
We resolve the crustal dynamics with $(n_{\rm max},l_{\rm max},m_{\rm max}) = (50,100,100)$, for a total of 1 million magneto-elastic eigenmodes.

The crustal dynamics of all simulations is qualitatively similar and weakly dependent on the initial quake location. 
The only difference with changing $\theta_Q$ is the angle of the external magnetic field with respect to the magnetar surface, and this is a relatively weak effect on the internal crustal dynamics. 
In all simulations the initial crustal perturbation launches magneto-elastic waves that propagate toward the surface and the crust-core interface. 
The waves bounce between the surface and the bottom of the crust with the characteristic frequency $\sim v_s/H \sim $~kHz, 
where $v_s =\sqrt{\mu/\rho}\approx 10^8$~cm~s$^{-1}$ in the deep crust and $H\sim 10^5$~cm is the thickness of the crust.
We also observe surface waves that propagate around the circumference of the neutron star. 
However, these waves contain a small fraction of the quake energy. 
Most of the seismic waves bounce radially and are strongly damped before they can spread horizontally around the crust. 
An example of the evolution for $\theta_Q = 3\pi /10$ is shown in Figure~\ref{quake}. 
Note that the locations of large $v_\theta$ correspond to fast mode emission, and the locations of large $v_\phi$ correspond to Alfv\'en wave emission.

The time evolution of different forms of energy is shown in Figure~\ref{fig:E/E_0alpha}. 
One can see that most of the crust quakes energy enters the core within $\tau_{\rm core}\sim 10$~ms, and a small fraction of the energy is converted into force-free waves in the magnetosphere.
We find that the energy in the crust decays nearly exponentially, according to 
\begin{equation}
E_{\rm crust}\approx E_Q {\rm exp}\left(-\frac{t}{\tau_{\rm core}}\right),
\label{decay}
\end{equation}
where $\tau_{\rm core}\sim 10 \ \rm ms$ is the decay time due to crust-core coupling from the analytic model (see Equation~(\ref{eq:tau_core})).

The fraction of energy emitted into the magnetosphere is clearly dependent on the initial quake location $\theta_Q$. 
For $\theta_Q = 0$ (polar cap quake), the field lines in the quake region are nearly radial $\theta_0 \approx 0$ and a larger fraction of the quake energy is extracted through the surface. 
For $\theta_Q = \pi/2$ (equatorial quake), the radial magnetic field in the quake region is weak and the transmission of energy into the magnetosphere is suppressed.

\begin{figure}
\includegraphics[width=0.45\textwidth]
{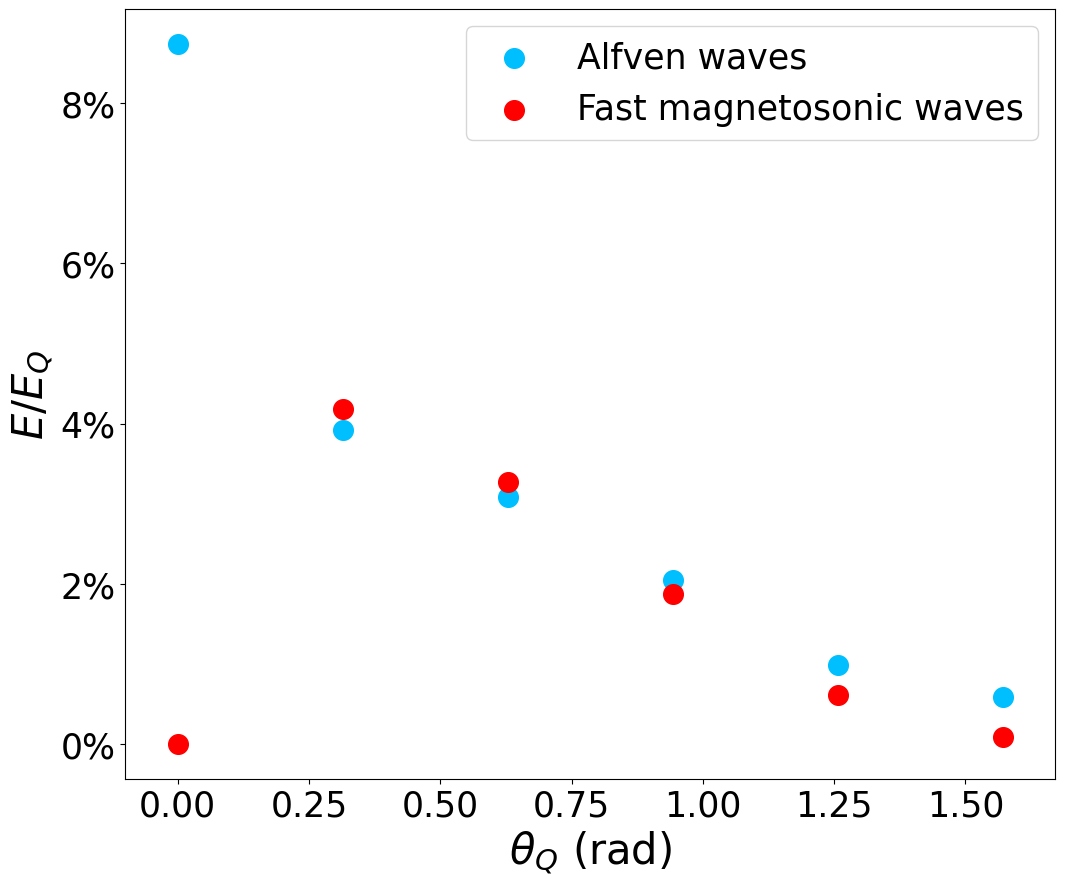}
\caption{The energy emitted into Alfv\'en waves (blue dot) and fast magnetosonic waves (red dot) as a function of polar angle $\theta_Q$ from $0-90^{\circ}$.}
\label{fig:E_FAratio}
\end{figure}

In Figure~\ref{fig:E_FAratio} we show the fraction of energy emitted into Alfv\'en and fast modes as a function of the initial quake location $\theta_Q$. In order to compare this with our analytic transmission calculations, we note that the total energy of the quake is shared equally between  strain energy and kinetic energy of the elastic waves. 
Therefore, the total energy of the quake may be written as the sum of two contributions $E_Q \approx E_{Q}^F + E_{Q}^A$, where  $E_{Q}^F$ is the kinetic energy associated with $\theta$ motions of the crust, and $E_{Q}^A$ is the kinetic energy associated with $\phi$ motions of the crust. The energy $E_Q^F$ is available to emit fast modes, and the energy $E_Q^A$ is available to emit Alfv\'en waves.

For the quake centered on the polar axis with $\theta_Q = 0$ the displacement is purely in the $\phi$ direction. Therefore, $E_{Q}^F = 0$ and $E_{Q}^A = E_Q$. 
The energy emitted into fast modes is then $E^F = 0$, and the energy emitted into Alfv\'en modes is $E^A = {\cal T}^A_{\rm mag} E^A_Q \sim 0.05 E_Q$, in reasonable agreement with the numerical simulation (Figures~\ref{fig:E/E_0alpha} and \ref{fig:E_FAratio}). Here ${\cal T}^A_{\rm mag}\approx 0.05$ is the analytic transmission coefficient (Equation~(\ref{TA_polar})) evaluated near the polar axis.

The quake centered on the equator ($\theta_Q = \pi/2$) is symmetric about $\theta = \pi/2$ and $\phi=0$. 
Therefore, half of the elastic energy is available to emit fast modes, and half is available to emit Alfv\'en waves: $E_{Q}^F = E_{Q}^A = 0.5E_Q$. 
The energy emitted into fast modes is then $E^F = \langle {\cal T}^F_{\rm mag}\rangle E_{Q}^F \sim 5\times 10^{-4}E_Q$, and the energy emitted into Alfv\'en waves is $E^A = \langle {\cal T}^A_{\rm mag} \rangle E^A_Q \sim 5\times 10^{-3}E_Q$, in reasonable agreement with the numerical simulations (Figures~\ref{fig:E/E_0alpha} and \ref{fig:E_FAratio}). 
Here $\langle {\cal T}^A_{\rm mag} \rangle \approx 10^{-2}$ and $\langle {\cal T}^F_{\rm mag} \rangle \approx 10^{-3}$ are the local spatial averages of Equations~(\ref{TA_eq}) and (\ref{TF_eq}) over the quake region. 
By averaging the transmission coefficients over the quake area, we account for the suppression of ${\cal T}^{A/F}_\text{mag}$ as $\theta\rightarrow \pi/2$.

\section{Discussion}\label{sec:Conclusion}

The accumulation of stress in magnetar crusts occurs due to the evolution of their internal magnetic fields, which has been studied in detail in previous works \citep{wood_three_2015, igoshev_2021, pons_2025, Bransgrove_2025b}. However, modeling of the elasto-plastic response of the crust is still in the early stages \citep{beloborodov_thermoplastic_2014, li_magnetar_2016, thompson_global_2017, Bransgrove_2018, lander_magnetic-field_2019, Gourgouliatos_2021, kojima2024, Bransgrove_2025}, and the yielding physics itself is also uncertain \citep{jones_nature_2003,Horowitz&Kadau2009}. In addition, special magnetic field configurations may be required to enable efficient excitation of elastic waves \citep{levin_dynamics_2012}. 
Despite these theoretical uncertainties, star quakes have been frequently invoked as a mechanism to trigger sudden transients in the magnetospheres of neutron stars \citep{Blaes1989,thompson_soft_1996,Bransgrove2020, Yuan2020, Beloborodov_2023}. 
In this work, we do not attempt to model the accumulation of stress, or the dynamics of a crustal failure. Instead, we focus on the propagation of elastic waves that 
may be excited by such an event, and how they are transmitted  into the magnetosphere.

We find that the internal crust dynamics is weakly affected by the quake location with respect to the magnetic axis. 
However, the location of the quake relative to the magnetic axis has a strong influence on the magnetospheric wave emission, since the external magnetic field determines the partition of energy into different magnetospheric wave modes. This dependence arises because the inclination of background magnetic field lines relative to the stellar surface controls the transmission coefficients into Alfv\'en and fast modes (see Section~\ref{sec:transmission_mag} for a detailed discussion).
For all initial conditions, the elastic waves bounce between the crust-core interface and the surface with the characteristic timescale $H/v_s \sim 1$~ms which gives the characteristic frequency of magnetospheric wave emission $\sim$~kHz. The quake is damped on a timescale $\tau_{\rm core}\sim 10~b^{-1/2} \ \rm ms$ as most of the energy is drained into the liquid core as Alfv\'en waves. We find that for all locations of the initial quake $\theta_Q$, a small fraction of the quake energy is transmitted into force-free waves in the magnetosphere $E^{A/F}/E_Q\sim 10^{-2} -10^{-1}$. Since the damping time of the quake is faster than $\pi R_{\star} / {v}_s \sim 30 \ \rm ms$, the waves are lost to the core before they can spread horizontally around the star. Therefore, most of the quake energy remains near the initial epicenter $\theta \approx \theta_Q$, and we expect this to be the site of the most powerful magnetospheric wave emission. We expect that the characteristic $\sim$~kHz frequency and decay on timescale $\tau_{\rm core}\sim 10~b^{-1/2} \ \rm ms$ may be observed as short-lived QPOs in magnetar X-ray bursts and hyperactive repeating FRBs \citep{ZhangJS2025,ZhouDK2025}. QPOs with similar frequencies and decay times were recently observed in a non-repeating FRB \citep{shuo_2026}.

Although our model does not track the escaping fast modes, recent magnetohydrodynamic (MHD) solutions of their propagation show that they can steepen and form monster radiative shocks in the outer magnetosphere \citep{Beloborodov_2023}. The shock formation was also demonstrated from first-principles with kinetic plasma simulations \citep{Chen_2022,bernardi2025globalkineticsimulationsmonster,Vanthieghem&Levinson2024}. 
The shocks accelerate plasma that rapidly cools by radiating high-energy photons, with re-processing of the radiation into an optically thick electron-positron pair plasma. Therefore, the fast modes emitted in our simulation should produce powerful X-ray bursts, with luminosity similar to that of the initial quake \citep{Beloborodov_2023}. 
As the shocks propagate, they develop into relativistic blast waves that expand adiabatically into the wind beyond the magnetosphere. 
The blast waves are expected to emit coherent radio waves, as demonstrated by first-principles kinetic simulations \citep{Plotnikov&Sironi2019,Sironi2021}. 
Inside the magnetar magnetosphere, fast modes can freely propagate through the open field line region, where high-frequency FRBs are expected to be produced via ICS off $\sim\rm kHz$ fast modes by charged bunches located several hundred magnetar radii away \citep{Zhang2022,Qu&Zhang2023,Qu&Zhang2024}. The frequency-dependent cross section predicts narrow spectra from incident high-frequency fast modes generated by crust quakes.
Future observations of FRBs will help test the predictions of each scenario.

The method we used for simulating magnetar quakes in 3D can be coupled with detailed simulations of the magnetosphere using force-free electrodynamics \citep{burnaz_2025}, relativistic MHD, and particle-in-cell methods \citep{bernardi2025globalkineticsimulationsmonster}. This will be essential to investigate the launching of magnetospheric waves by more general crustal deformations, as well as the formation of magnetospheric shocks that could produce fast radio burst emission. Also note that we simplified the crust-core coupling by treating the core as a perfect reservoir for Alfv\'en waves. A more detailed model would include the self-consistent dynamical evolution of the core and its feedback on the crust, although this is unlikely to change our conclusions regarding emission of fast modes from the surface at high frequencies.

\section*{Acknowledgments}
We thank Yuri Levin, Andrei Beloborodov, and Chris Thompson for helpful discussions and an
anonymous referee for helpful comments.
Y.Q. is supported by the Top Tier Doctoral Graduate Research Assistantship (TTDGRA) at University of Nevada, Las Vegas. A.B. is supported by a PCTS fellowship and a Lyman Spitzer Jr. fellowship.


\begin{thebibliography}{}
\expandafter\ifx\csname natexlab\endcsname\relax\def\natexlab#1{#1}\fi
\providecommand{\url}[1]{\href{#1}{#1}}
\providecommand{\dodoi}[1]{doi:~\href{http://doi.org/#1}{\nolinkurl{#1}}}
\providecommand{\doeprint}[1]{\href{http://ascl.net/#1}{\nolinkurl{http://ascl.net/#1}}}
\providecommand{\doarXiv}[1]{\href{https://arxiv.org/abs/#1}{\nolinkurl{https://arxiv.org/abs/#1}}}

\bibitem[{Akgün {et~al.}(2017)Akgün, Cerdá–Durán, Miralles, \& Pons}]{Akgun_2017}
Akgün, T., Cerdá–Durán, P., Miralles, J.~A., \& Pons, J.~A. 2017, Monthly Notices of the Royal Astronomical Society, 472, 3914–3923, \dodoi{10.1093/mnras/stx2235}

\bibitem[{Baym {et~al.}(1969)Baym, Pethick, \& Pines}]{baym_superfluidity_1969}
Baym, G., Pethick, C., \& Pines, D. 1969, Nature, 224, 673, \dodoi{10.1038/224673a0}

\bibitem[{{Beloborodov}(2017)}]{Beloborodov2017}
{Beloborodov}, A.~M. 2017, \apjl, 843, L26, \dodoi{10.3847/2041-8213/aa78f3}

\bibitem[{{Beloborodov}(2020)}]{Beloborodov2020}
---. 2020, \apj, 896, 142, \dodoi{10.3847/1538-4357/ab83eb}

\bibitem[{{Beloborodov}(2023)}]{Beloborodov_2023}
---. 2023, \apj, 959, 34, \dodoi{10.3847/1538-4357/acf659}

\bibitem[{Beloborodov \& Levin(2014)}]{beloborodov_thermoplastic_2014}
Beloborodov, A.~M., \& Levin, Y. 2014, The Astrophysical Journal, 794, L24, \dodoi{10.1088/2041-8205/794/2/L24}

\bibitem[{Bernardi {et~al.}(2025)Bernardi, Yuan, \& Chen}]{bernardi2025globalkineticsimulationsmonster}
Bernardi, D., Yuan, Y., \& Chen, A.~Y. 2025, Global Kinetic Simulations of Monster Shocks and Their Emission.
\newblock \doarXiv{2506.04175}

\bibitem[{{Blaes} {et~al.}(1989){Blaes}, {Blandford}, {Goldreich}, \& {Madau}}]{Blaes1989}
{Blaes}, O., {Blandford}, R., {Goldreich}, P., \& {Madau}, P. 1989, \apj, 343, 839, \dodoi{10.1086/167754}

\bibitem[{{Bochenek} {et~al.}(2020){Bochenek}, {Ravi}, {Belov}, {Hallinan}, {Kocz}, {Kulkarni}, \& {McKenna}}]{Bochenek2020}
{Bochenek}, C.~D., {Ravi}, V., {Belov}, K.~V., {et~al.} 2020, \nat, 587, 59, \dodoi{10.1038/s41586-020-2872-x}

\bibitem[{{Bransgrove} {et~al.}(2020){Bransgrove}, {Beloborodov}, \& {Levin}}]{Bransgrove2020}
{Bransgrove}, A., {Beloborodov}, A.~M., \& {Levin}, Y. 2020, \apj, 897, 173, \dodoi{10.3847/1538-4357/ab93b7}

\bibitem[{{Bransgrove} {et~al.}(2025{\natexlab{a}}){Bransgrove}, {Beloborodov}, \& {Levin}}]{Bransgrove_2025b}
---. 2025{\natexlab{a}}, arXiv e-prints, arXiv:2508.13419, \dodoi{10.48550/arXiv.2508.13419}

\bibitem[{{Bransgrove} {et~al.}(2018){Bransgrove}, {Levin}, \& {Beloborodov}}]{Bransgrove_2018}
{Bransgrove}, A., {Levin}, Y., \& {Beloborodov}, A. 2018, \mnras, 473, 2771, \dodoi{10.1093/mnras/stx2508}

\bibitem[{{Bransgrove} {et~al.}(2025{\natexlab{b}}){Bransgrove}, {Levin}, \& {Beloborodov}}]{Bransgrove_2025}
{Bransgrove}, A., {Levin}, Y., \& {Beloborodov}, A.~M. 2025{\natexlab{b}}, \apj, 979, 144, \dodoi{10.3847/1538-4357/ad90a3}

\bibitem[{{Burnaz} {et~al.}(2025){Burnaz}, {Most}, \& {Bransgrove}}]{burnaz_2025}
{Burnaz}, L., {Most}, E.~R., \& {Bransgrove}, A. 2025, arXiv e-prints, arXiv:2508.18033, \dodoi{10.48550/arXiv.2508.18033}

\bibitem[{Caplan {et~al.}(2018)Caplan, Schneider, \& Horowitz}]{Caplan_2018}
Caplan, M., Schneider, A., \& Horowitz, C. 2018, Physical Review Letters, 121, \dodoi{10.1103/physrevlett.121.132701}

\bibitem[{{Chen} {et~al.}(2022){Chen}, {Yuan}, {Li}, \& {Mahlmann}}]{Chen_2022}
{Chen}, A.~Y., {Yuan}, Y., {Li}, X., \& {Mahlmann}, J.~F. 2022, arXiv e-prints, arXiv:2210.13506, \dodoi{10.48550/arXiv.2210.13506}

\bibitem[{{CHIME/FRB Collaboration} {et~al.}(2020){CHIME/FRB Collaboration}, {Andersen}, {Bandura}, {Bhardwaj}, {Bij}, {Boyce}, {Boyle}, {Brar}, {Cassanelli}, {Chawla}, {Chen}, {Cliche}, {Cook}, {Cubranic}, {Curtin}, {Denman}, {Dobbs}, {Dong}, {Fandino}, {Fonseca}, {Gaensler}, {Giri}, {Good}, {Halpern}, {Hill}, {Hinshaw}, {H{\"o}fer}, {Josephy}, {Kania}, {Kaspi}, {Landecker}, {Leung}, {Li}, {Lin}, {Masui}, {McKinven}, {Mena-Parra}, {Merryfield}, {Meyers}, {Michilli}, {Milutinovic}, {Mirhosseini}, {M{\"u}nchmeyer}, {Naidu}, {Newburgh}, {Ng}, {Patel}, {Pen}, {Pinsonneault-Marotte}, {Pleunis}, {Quine}, {Rafiei-Ravandi}, {Rahman}, {Ransom}, {Renard}, {Sanghavi}, {Scholz}, {Shaw}, {Shin}, {Siegel}, {Singh}, {Smegal}, {Smith}, {Stairs}, {Tan}, {Tendulkar}, {Tretyakov}, {Vanderlinde}, {Wang}, {Wulf}, \& {Zwaniga}}]{CHIME/FRB2020}
{CHIME/FRB Collaboration}, {Andersen}, B.~C., {Bandura}, K.~M., {et~al.} 2020, \nat, 587, 54, \dodoi{10.1038/s41586-020-2863-y}

\bibitem[{Douchin \& Haensel(2001)}]{douchin_unified_2001}
Douchin, F., \& Haensel, P. 2001, A\&A, 380, 151, \dodoi{10.1051/0004-6361:20011402}

\bibitem[{Easson \& Pethick(1977)}]{easson_stress_1977}
Easson, I., \& Pethick, C.~J. 1977, Phys. Rev. D, 16, 275, \dodoi{10.1103/PhysRevD.16.275}

\bibitem[{Gabler {et~al.}(2014)Gabler, Cerdá-Durán, Stergioulas, Font, \& Müller}]{Gabler_2014}
Gabler, M., Cerdá-Durán, P., Stergioulas, N., Font, J.~A., \& Müller, E. 2014, Monthly Notices of the Royal Astronomical Society, 443, 1416–1424, \dodoi{10.1093/mnras/stu1263}

\bibitem[{Gourgouliatos \& Lander(2021)}]{Gourgouliatos_2021}
Gourgouliatos, K.~N., \& Lander, S.~K. 2021, Monthly Notices of the Royal Astronomical Society, 506, 3578–3587, \dodoi{10.1093/mnras/stab1869}

\bibitem[{Haensel \& Potekhin(2004)}]{haensel_analytical_2004}
Haensel, P., \& Potekhin, A.~Y. 2004, A\&A, 428, 191, \dodoi{10.1051/0004-6361:20041722}

\bibitem[{{Harding} \& {Lai}(2006)}]{Harding&Lai2006}
{Harding}, A.~K., \& {Lai}, D. 2006, Reports on Progress in Physics, 69, 2631, \dodoi{10.1088/0034-4885/69/9/R03}

\bibitem[{{Horowitz} \& {Kadau}(2009)}]{Horowitz&Kadau2009}
{Horowitz}, C.~J., \& {Kadau}, K. 2009, \prl, 102, 191102, \dodoi{10.1103/PhysRevLett.102.191102}

\bibitem[{{Hurley} {et~al.}(2005){Hurley}, {Boggs}, {Smith}, {Duncan}, {Lin}, {Zoglauer}, {Krucker}, {Hurford}, {Hudson}, {Wigger}, {Hajdas}, {Thompson}, {Mitrofanov}, {Sanin}, {Boynton}, {Fellows}, {von Kienlin}, {Lichti}, {Rau}, \& {Cline}}]{Hurley_2005}
{Hurley}, K., {Boggs}, S.~E., {Smith}, D.~M., {et~al.} 2005, \nat, 434, 1098, \dodoi{10.1038/nature03519}

\bibitem[{Igoshev {et~al.}(2021)Igoshev, Popov, \& Hollerbach}]{igoshev_2021}
Igoshev, A.~P., Popov, S.~B., \& Hollerbach, R. 2021, Evolution of neutron star magnetic fields.
\newblock \doarXiv{2109.05584}

\bibitem[{Jones(2003)}]{jones_nature_2003}
Jones, P.~B. 2003, ApJ, 595, 342, \dodoi{10.1086/377351}

\bibitem[{{Kaspi} \& {Beloborodov}(2017)}]{Kaspi&Beloborodov2017}
{Kaspi}, V.~M., \& {Beloborodov}, A.~M. 2017, \araa, 55, 261, \dodoi{10.1146/annurev-astro-081915-023329}

\bibitem[{Kojima(2024)}]{kojima2024}
Kojima, Y. 2024, Correct criterion of crustal failure driven by intense magnetic stress in neutron stars.
\newblock \doarXiv{2408.14100}

\bibitem[{Komissarov(2004)}]{komissarov_electrodynamics_2004}
Komissarov, S.~S. 2004, Mon Not R Astron Soc, 350, 427, \dodoi{10.1111/j.1365-2966.2004.07598.x}

\bibitem[{{Kumar} \& {Bo{\v{s}}njak}(2020)}]{Kumar&Bosnjak2020}
{Kumar}, P., \& {Bo{\v{s}}njak}, {\v{Z}}. 2020, \mnras, 494, 2385, \dodoi{10.1093/mnras/staa774}

\bibitem[{{Kumar} {et~al.}(2022){Kumar}, {Gill}, \& {Lu}}]{Kumar2022}
{Kumar}, P., {Gill}, R., \& {Lu}, W. 2022, \mnras, 516, 2697, \dodoi{10.1093/mnras/stac2446}

\bibitem[{{Kumar} {et~al.}(2017){Kumar}, {Lu}, \& {Bhattacharya}}]{Kumar2017}
{Kumar}, P., {Lu}, W., \& {Bhattacharya}, M. 2017, \mnras, 468, 2726, \dodoi{10.1093/mnras/stx665}

\bibitem[{{Landau} \& {Lifshitz}(1959)}]{Landau1959}
{Landau}, L.~D., \& {Lifshitz}, E.~M. 1959, {Theory of elasticity}

\bibitem[{Lander \& Gourgouliatos(2019)}]{lander_magnetic-field_2019}
Lander, S.~K., \& Gourgouliatos, K.~N. 2019, arXiv:1902.02121 [astro-ph].
\newblock \url{http://arxiv.org/abs/1902.02121}

\bibitem[{Levin(2006)}]{levin_qpos_2006}
Levin, Y. 2006, MNRASL, 368, L35, \dodoi{10.1111/j.1745-3933.2006.00155.x}

\bibitem[{Levin \& Lyutikov(2012)}]{levin_dynamics_2012}
Levin, Y., \& Lyutikov, M. 2012, MNRAS, 427, 1574, \dodoi{10.1111/j.1365-2966.2012.22016.x}

\bibitem[{{Li} {et~al.}(2021){Li}, {Lin}, {Xiong}, {Ge}, {Li}, {Li}, {Lu}, {Zhang}, {Tuo}, {Nang}, {Zhang}, {Xiao}, {Chen}, {Song}, {Xu}, {Liu}, {Jia}, {Cao}, {Qu}, {Zhang}, {Gu}, {Liao}, {Zhao}, {Tan}, {Nie}, {Zhao}, {Zheng}, {Zheng}, {Luo}, {Cai}, {Li}, {Xue}, {Bu}, {Chang}, {Chen}, {Chen}, {Chen}, {Chen}, {Chen}, {Cui}, {Cui}, {Deng}, {Dong}, {Du}, {Fu}, {Gao}, {Gao}, {Gao}, {Gu}, {Guan}, {Guo}, {Han}, {Huang}, {Huo}, {Jiang}, {Jiang}, {Jin}, {Jin}, {Kong}, {Li}, {Li}, {Li}, {Li}, {Li}, {Li}, {Li}, {Liang}, {Liu}, {Liu}, {Liu}, {Liu}, {Liu}, {Lu}, {Lu}, {Luo}, {Ma}, {Meng}, {Ou}, {Sai}, {Shang}, {Song}, {Sun}, {Tao}, {Wang}, {Wang}, {Wang}, {Wang}, {Wang}, {Wen}, {Wu}, {Wu}, {Wu}, {Xiao}, {Xu}, {Yang}, {Yang}, {Yang}, {Yang}, {Yi}, {Yin}, {You}, {Zhang}, {Zhang}, {Zhang}, {Zhang}, {Zhang}, {Zhang}, {Zhang}, {Zhang}, {Zhang}, {Zhou}, {Zhou}, {Zhu}, {Zhu}, \& {Zhuang}}]{CKLi21}
{Li}, C.~K., {Lin}, L., {Xiong}, S.~L., {et~al.} 2021, Nature Astronomy, 5, 378.
\newblock \doarXiv{2005.11071}

\bibitem[{{Li} \& {Beloborodov}(2015)}]{Li&Beloborodov2015}
{Li}, X., \& {Beloborodov}, A.~M. 2015, \apj, 815, 25, \dodoi{10.1088/0004-637X/815/1/25}

\bibitem[{Li {et~al.}(2016)Li, Levin, \& Beloborodov}]{li_magnetar_2016}
Li, X., Levin, Y., \& Beloborodov, A.~M. 2016, arXiv:1606.04895 [astro-ph].
\newblock \url{http://arxiv.org/abs/1606.04895}

\bibitem[{{Lu} {et~al.}(2020){Lu}, {Kumar}, \& {Zhang}}]{Lu2020}
{Lu}, W., {Kumar}, P., \& {Zhang}, B. 2020, \mnras, 498, 1397, \dodoi{10.1093/mnras/staa2450}

\bibitem[{{Lyubarsky}(2014)}]{Lyubarsky2014}
{Lyubarsky}, Y. 2014, \mnras, 442, L9, \dodoi{10.1093/mnrasl/slu046}

\bibitem[{{Lyubarsky}(2021)}]{Lyubarsky2021}
---. 2021, Universe, 7, 56, \dodoi{10.3390/universe7030056}

\bibitem[{{Lyutikov}(2021)}]{Lyutikov2021}
{Lyutikov}, M. 2021, \apj, 922, 166, \dodoi{10.3847/1538-4357/ac1b32}

\bibitem[{{Lyutikov} \& {Freund}(2025)}]{Lyutikov&Freund2025}
{Lyutikov}, M., \& {Freund}, H. 2025, Journal of Plasma Physics, 91, E10, \dodoi{10.1017/S0022377824000497}

\bibitem[{{Mahlmann} {et~al.}(2022){Mahlmann}, {Philippov}, {Levinson}, {Spitkovsky}, \& {Hakobyan}}]{Mahlmann2022}
{Mahlmann}, J.~F., {Philippov}, A.~A., {Levinson}, A., {Spitkovsky}, A., \& {Hakobyan}, H. 2022, \apjl, 932, L20, \dodoi{10.3847/2041-8213/ac7156}

\bibitem[{{Mereghetti}(2008)}]{Mereghetti2008}
{Mereghetti}, S. 2008, \aapr, 15, 225, \dodoi{10.1007/s00159-008-0011-z}

\bibitem[{{Mereghetti} {et~al.}(2020){Mereghetti}, {Savchenko}, {Ferrigno}, {G{\"o}tz}, {Rigoselli}, {Tiengo}, {Bazzano}, {Bozzo}, {Coleiro}, {Courvoisier}, {Doyle}, {Goldwurm}, {Hanlon}, {Jourdain}, {von Kienlin}, {Lutovinov}, {Martin-Carrillo}, {Molkov}, {Natalucci}, {Onori}, {Panessa}, {Rodi}, {Rodriguez}, {S{\'a}nchez-Fern{\'a}ndez}, {Sunyaev}, \& {Ubertini}}]{Mereghetti20}
{Mereghetti}, S., {Savchenko}, V., {Ferrigno}, C., {et~al.} 2020, \apjl, 898, L29, \dodoi{10.3847/2041-8213/aba2cf}

\bibitem[{{Metzger} {et~al.}(2019){Metzger}, {Margalit}, \& {Sironi}}]{Metzger2019}
{Metzger}, B.~D., {Margalit}, B., \& {Sironi}, L. 2019, \mnras, 485, 4091, \dodoi{10.1093/mnras/stz700}

\bibitem[{{Palmer} {et~al.}(2005){Palmer}, {Barthelmy}, {Gehrels}, {Kippen}, {Cayton}, {Kouveliotou}, {Eichler}, {Wijers}, {Woods}, {Granot}, {Lyubarsky}, {Ramirez-Ruiz}, {Barbier}, {Chester}, {Cummings}, {Fenimore}, {Finger}, {Gaensler}, {Hullinger}, {Krimm}, {Markwardt}, {Nousek}, {Parsons}, {Patel}, {Sakamoto}, {Sato}, {Suzuki}, \& {Tueller}}]{Palmer_2005}
{Palmer}, D.~M., {Barthelmy}, S., {Gehrels}, N., {et~al.} 2005, \nat, 434, 1107, \dodoi{10.1038/nature03525}

\bibitem[{{Plotnikov} \& {Sironi}(2019)}]{Plotnikov&Sironi2019}
{Plotnikov}, I., \& {Sironi}, L. 2019, \mnras, 485, 3816, \dodoi{10.1093/mnras/stz640}

\bibitem[{Pons {et~al.}(2025)Pons, Dehman, \& Viganò}]{pons_2025}
Pons, J.~A., Dehman, C., \& Viganò, D. 2025, Magnetic, thermal and rotational evolution of isolated neutron stars.
\newblock \doarXiv{2509.06699}

\bibitem[{{Potekhin} \& {Chabrier}(2000)}]{Potekhin&Chabrier2000}
{Potekhin}, A.~Y., \& {Chabrier}, G. 2000, \pre, 62, 8554, \dodoi{10.1103/PhysRevE.62.8554}

\bibitem[{{Qu} \& {Zhang}(2023)}]{Qu&Zhang2023}
{Qu}, Y., \& {Zhang}, B. 2023, \mnras, 522, 2448, \dodoi{10.1093/mnras/stad1072}

\bibitem[{{Qu} \& {Zhang}(2024)}]{Qu&Zhang2024}
---. 2024, \apj, 972, 124, \dodoi{10.3847/1538-4357/ad5d5b}

\bibitem[{{Ridnaia} {et~al.}(2021){Ridnaia}, {Svinkin}, {Frederiks}, {Bykov}, {Popov}, {Aptekar}, {Golenetskii}, {Lysenko}, {Tsvetkova}, {Ulanov}, \& {Cline}}]{konus}
{Ridnaia}, A., {Svinkin}, D., {Frederiks}, D., {et~al.} 2021, Nature Astronomy, 5, 372, \dodoi{10.1038/s41550-020-01265-0}

\bibitem[{{Ruderman}(1968)}]{Ruderman1968}
{Ruderman}, M.~A. 1968, \nat, 218, 1128, \dodoi{10.1038/2181128a0}

\bibitem[{{Sironi} {et~al.}(2021){Sironi}, {Plotnikov}, {N{\"a}ttil{\"a}}, \& {Beloborodov}}]{Sironi2021}
{Sironi}, L., {Plotnikov}, I., {N{\"a}ttil{\"a}}, J., \& {Beloborodov}, A.~M. 2021, \prl, 127, 035101, \dodoi{10.1103/PhysRevLett.127.035101}

\bibitem[{{Strohmayer} {et~al.}(1991){Strohmayer}, {Ogata}, {Iyetomi}, {Ichimaru}, \& {van Horn}}]{Strohmayer1991}
{Strohmayer}, T., {Ogata}, S., {Iyetomi}, H., {Ichimaru}, S., \& {van Horn}, H.~M. 1991, \apj, 375, 679, \dodoi{10.1086/170231}

\bibitem[{{Tavani} {et~al.}(2021){Tavani}, {Casentini}, {Ursi}, {Verrecchia}, {Addis}, {Antonelli}, {Argan}, {Barbiellini}, {Baroncelli}, {Bernardi}, {Bianchi}, {Bulgarelli}, {Caraveo}, {Cardillo}, {Cattaneo}, {Chen}, {Costa}, {Del Monte}, {Di Cocco}, {Di Persio}, {Donnarumma}, {Evangelista}, {Feroci}, {Ferrari}, {Fioretti}, {Fuschino}, {Galli}, {Gianotti}, {Giuliani}, {Labanti}, {Lazzarotto}, {Lipari}, {Longo}, {Lucarelli}, {Magro}, {Marisaldi}, {Mereghetti}, {Morelli}, {Morselli}, {Naldi}, {Pacciani}, {Parmiggiani}, {Paoletti}, {Pellizzoni}, {Perri}, {Perotti}, {Piano}, {Picozza}, {Pilia}, {Pittori}, {Puccetti}, {Pupillo}, {Rapisarda}, {Rappoldi}, {Rubini}, {Setti}, {Soffitta}, {Trifoglio}, {Trois}, {Vercellone}, {Vittorini}, {Giommi}, \& {D'Amico}}]{AGILE}
{Tavani}, M., {Casentini}, C., {Ursi}, A., {et~al.} 2021, Nature Astronomy, 5, 401, \dodoi{10.1038/s41550-020-01276-x}

\bibitem[{Thompson \& Duncan(1995)}]{thompson_soft_1995}
Thompson, C., \& Duncan, R.~C. 1995, MNRAS, 275, 255, \dodoi{10.1093/mnras/275.2.255}

\bibitem[{{Thompson} \& Duncan(1996)}]{thompson_soft_1996}
{Thompson}, C., \& Duncan, R.~C. 1996, The Astrophysical Journal, 473, 322, \dodoi{10.1086/178147}

\bibitem[{Thompson {et~al.}(2017)Thompson, Yang, \& Ortiz}]{thompson_global_2017}
Thompson, C., Yang, H., \& Ortiz, N. 2017, ApJ, 841, 54, \dodoi{10.3847/1538-4357/aa6c30}

\bibitem[{van Hoven \& Levin(2008)}]{van_hoven_hydromagnetic_2008}
van Hoven, M., \& Levin, Y. 2008, MNRAS, 391, 283, \dodoi{10.1111/j.1365-2966.2008.13881.x}

\bibitem[{{Vanthieghem} \& {Levinson}(2024)}]{Vanthieghem&Levinson2024}
{Vanthieghem}, A., \& {Levinson}, A. 2024, arXiv e-prints, arXiv:2407.15076, \dodoi{10.48550/arXiv.2407.15076}

\bibitem[{Wood \& Hollerbach(2015)}]{wood_three_2015}
Wood, T.~S., \& Hollerbach, R. 2015, Phys. Rev. Lett., 114, 191101, \dodoi{10.1103/PhysRevLett.114.191101}

\bibitem[{{Woods} \& {Thompson}(2006)}]{Woods&Thompson2006}
{Woods}, P.~M., \& {Thompson}, C. 2006, in Compact stellar X-ray sources, ed. W.~H.~G. {Lewin} \& M.~{van der Klis}, Vol.~39, 547--586, \dodoi{10.48550/arXiv.astro-ph/0406133}

\bibitem[{{Xiao} {et~al.}(2026){Xiao}, {Jiang}, \& {Li}}]{shuo_2026}
{Xiao}, S., {Jiang}, Z.-H., \& {Li}, D. 2026, arXiv e-prints, arXiv:2601.03950, \dodoi{10.48550/arXiv.2601.03950}

\bibitem[{{Yuan} {et~al.}(2020){Yuan}, {Beloborodov}, {Chen}, \& {Levin}}]{Yuan2020}
{Yuan}, Y., {Beloborodov}, A.~M., {Chen}, A.~Y., \& {Levin}, Y. 2020, \apjl, 900, L21, \dodoi{10.3847/2041-8213/abafa8}

\bibitem[{{Yuan} {et~al.}(2022){Yuan}, {Beloborodov}, {Chen}, {Levin}, {Most}, \& {Philippov}}]{Yuan2022}
{Yuan}, Y., {Beloborodov}, A.~M., {Chen}, A.~Y., {et~al.} 2022, \apj, 933, 174, \dodoi{10.3847/1538-4357/ac7529}

\bibitem[{{Yuan} {et~al.}(2021){Yuan}, {Levin}, {Bransgrove}, \& {Philippov}}]{Yuan2021}
{Yuan}, Y., {Levin}, Y., {Bransgrove}, A., \& {Philippov}, A. 2021, \apj, 908, 176, \dodoi{10.3847/1538-4357/abd405}

\bibitem[{{Zhang}(2020)}]{Zhang2020}
{Zhang}, B. 2020, \nat, 587, 45, \dodoi{10.1038/s41586-020-2828-1}

\bibitem[{{Zhang}(2022)}]{Zhang2022}
---. 2022, \apj, 925, 53, \dodoi{10.3847/1538-4357/ac3979}

\bibitem[{{Zhang}(2023)}]{ZhangRMP}
---. 2023, Reviews of Modern Physics, 95, 035005, \dodoi{10.1103/RevModPhys.95.035005}

\bibitem[{{Zhang} {et~al.}(2025){Zhang}, {Wang}, {Wang}, {Wu}, {Li}, {Zhu}, {Zhang}, {Gao}, {Lee}, {Han}, {Tsai}, {Wang}, {Huang}, {Zou}, {Zhou}, {Lu}, {Xie}, {Fang}, {Cao}, {Miao}, {Zhu}, {Chen}, {Cheng}, {Ke}, {Zhang}, {Zhang}, {Cao}, {Tian}, {Wu}, {Zhang}, {Niu}, {Zhou}, {Xu}, {Wang}, {Chen}, {Chen}, {Cui}, {Feng}, {G{\"u}gercino{\u{g}}lu}, {Huang}, {Li}, {Li}, {Li}, {Lin}, {Liu}, {Luo}, {Luo}, {Niu}, {Qu}, {Qu}, {Menberu Tedila}, {Wang}, {Wang}, {Wang}, {Wang}, {Weng}, {Wu}, {Xu}, {Yang}, {Yang}, {Yew}, {Yu}, {Zhang}, \& {Zhao}}]{ZhangJS2025}
{Zhang}, J.-S., {Wang}, T.-C., {Wang}, P., {et~al.} 2025, arXiv e-prints, arXiv:2507.14707, \dodoi{10.48550/arXiv.2507.14707}

\bibitem[{{Zhou} {et~al.}(2025){Zhou}, {Wang}, {Fang}, {Zhu}, {Zhang}, {Li}, {Feng}, {Huang}, {Lee}, {Han}, {Zou}, {Zhang}, {Xiao}, {Luo}, {Zhang}, {Wang}, {Lu}, {Cao}, {Yu}, {Li}, {Miao}, {Xie}, {Chen}, {Wang}, {Qu}, {Chen}, {Zhu}, {Cao}, {Chen}, {Du}, {Gao}, {Huang}, {Li}, {Li}, {Li}, {Lin}, {Liu}, {Luo}, {Niu}, {Niu}, {Qu}, {Tian}, {Tsai}, {Wang}, {Wang}, {Wang}, {Wang}, {Weng}, {Wu}, {Wu}, {Xu}, {Yang}, {Yang}, {Yew}, {Zhang}, {Zhang}, {Zhang}, {Zhao}, \& {Zhou}}]{ZhouDK2025}
{Zhou}, D., {Wang}, P., {Fang}, J., {et~al.} 2025, arXiv e-prints, arXiv:2507.14708, \dodoi{10.48550/arXiv.2507.14708}

\end{thebibliography}

\end{document}